\pdfoutput=1
\documentclass[journal=accacs,manuscript=article,layout=twocolumn]{achemso}
\usepackage[version=4]{mhchem} % Formula subscripts using \ce{}

\newcommand\tabcaption{\def\@captype{table}\caption}
\newcommand\figcaption{\def\@captype{figure}\caption}
\usepackage{amsmath}
\usepackage{chemformula}    % \ch command for chemical formulae
\usepackage{bm}
\usepackage{multirow}
\usepackage{mdframed}
\usepackage{rotating}
\usepackage{booktabs}
\usepackage{etoolbox}
\usepackage{enumitem}
\usepackage{subcaption}
\usepackage[nonumberlist]{glossaries}
\mciteErrorOnUnknownfalse
\usepackage{xcolor}
\usepackage{comment}
\usepackage{array}
\usepackage[hidelinks]{hyperref}
\newcounter{magicrownumbers}
\preto\tabular{\setcounter{magicrownumbers}{0}}
\usepackage[textsize=small]{todonotes}
\usepackage{siunitx}
\usepackage{amssymb}
\usepackage{pifont}
\usepackage{threeparttable}
\usepackage{graphicx}

\newcommand{\dataset}[1]{ODAC23}

\newcommand{\carbondioxide}[0]{\ce{CO2}}
\newcommand{\water}[0]{\ce{H2O}}
\newcommand{\ev}{\text{eV}}
\newcommand{\angstrom}{\text{\normalfont\AA}}
\newcommand{\meva}{\text{m}\ev/\angstrom}
\newcommand{\eva}{\ev/\angstrom}
\newcommand{\emae}{Energy MAE [eV] $\downarrow$}
\newcommand{\fmae}{Force MAE [meV/\angstrom] $\downarrow$}
\newcommand{\fcos}{Force Cos $\uparrow$}
\newcommand{\efwt}{EFwT $\uparrow$}

\newcommand{\testid}{test-id}
\newcommand{\oodbig}{test-ood (big)}
\newcommand{\oodb}{test-ood(b)}
\newcommand{\oodlinker}{test-ood (linker)}
\newcommand{\oodl}{test-ood(l)}
\newcommand{\oodtopology}{test-ood (topology)}
\newcommand{\oodt}{test-ood(t)}
\newcommand{\oodlt}{test-ood(lt)}
\newcommand{\oodlinkertopo}{test-ood (linker \& topology)}
\newcommand{\cmark}{\ding{51}}
\newcommand{\xmark}{\ding{55}}

\newcommand{\npristine}{5,079}
\newcommand{\ndefective}{3,628}

\newacronym{DFT}{DFT}{Density Functional Theory}
\newacronym{SI}{SI}{Supporting Information}
\newacronym{PES}{PES}{Potential Energy Surface}
\newacronym{QE}{QE}{Quantum Espresso}
\newacronym{VASP}{VASP}{Vienna Ab initio Simulation Package}
\newacronym{S2EF}{\textit{S2EF}}{Structure to Energy and Forces}
\newacronym{RS2EF}{\textit{S2EF-Total}}{Structure to Total Energy and Forces}
\newacronym{RIS2RE}{\textit{IS2RE-Total}}{Initial Structure to Total Relaxed Energy}
\newacronym{IS2RE}{\textit{IS2RE}}{Initial Structure to Relaxed Energy}
\newacronym{IS2RS}{\textit{IS2RS}}{Initial Structure to Relaxed Structure}
\newacronym{ID}{ID}{In-Domain}
\newacronym{OOD}{OOD}{Out-of-Domain}
\newacronym{OC20}{OC20}{Open Catalyst 2020 Dataset}
\newacronym{ML}{ML}{machine learning}
\newacronym{GNNs}{GNNs}{Graph Neural Networks}
\newacronym{GNN}{GNN}{Graph Neural Network}
\newacronym{MD}{MD}{\textit{ab initio} Molecular Dynamics}
\newacronym{OER}{OER}{Oxygen Evolution Reaction}
\newacronym{MAE}{MAE}{Mean Absolute Error}

\newacronym{EFwT}{EFwT}{Energy and Forces within Threshold}
\newacronym{EwT}{EwT}{Energy within Threshold}
\newacronym{ADwT}{ADwT}{Average Distance within Threshold}
\newacronym{FbT}{FbT}{Force below Threshold}
\newacronym{AFbT}{AFbT}{Average Force below Threshold}
\newacronym{ODAC23}{ODAC23}{Open DAC 2023}
\newacronym{MvK}{MvK}{Mars-van Krevelen}
\usepackage[version=4]{mhchem}

\glsaddall

\newcommand{\fair}{Fundamental AI Research, Meta AI, Meta, Menlo Park, CA, USA}

\newcommand{\gatech}{School of Chemical and Biomolecular Engineering, Georgia Institute of Technology, Atlanta, GA, USA}
\newcommand{\ornl}{Oak Ridge National Laboratory, Oak Ridge, TN, USA}

\newcommand{\Npristine}{4,942}
\newcommand{\Ndefective}{3,470}

\author{Anuroop Sriram}
\affiliation{\fair}
\email{anuroops@meta.com}
\author{Sihoon Choi}
\affiliation{\fair}
\alsoaffiliation{\gatech}
\author{Xiaohan Yu}
\affiliation{\gatech}
\author{Logan M. Brabson}
\affiliation{\gatech}
\author{Abhishek Das}
\affiliation{\fair}
\author{Zachary Ulissi}
\affiliation{\fair}
\author{Matt Uyttendaele}
\affiliation{\fair}
\author{Andrew J. Medford}
\affiliation{\gatech}
\email{ajm@gatech.edu}
\author{David S. Sholl}
\affiliation{\gatech}
\alsoaffiliation{\ornl}
\email{shollds@ornl.gov}

\title[]
  {The Open DAC 2023 Dataset and Challenges for Sorbent Discovery in Direct Air Capture}

\abbreviations{IR,NMR,UV}
\keywords{Direct air capture, metal organic frameworks, carbon capture, density functional theory, datasets, machine learning, graph convolutions, force field}

\let\oldmaketitle\maketitle
\let\maketitle\relax

\begin{document}



\twocolumn[
\begin{@twocolumnfalse}
\oldmaketitle
\begin{abstract}
New methods for carbon dioxide removal are urgently needed to combat global climate change. Direct air capture (DAC) is an emerging technology to capture carbon dioxide directly from ambient air. Metal-organic frameworks (MOFs) have been widely studied as potentially customizable adsorbents for DAC. However, discovering promising MOF sorbents for DAC is challenging because of the vast chemical space to explore and the need to understand materials as functions of humidity and temperature. We explore a computational approach benefiting from recent innovations in machine learning (ML) and present a dataset named Open DAC 2023 (ODAC23) consisting of more than 38M density functional theory (DFT) calculations on more than 8,400 MOF materials containing adsorbed \ce{CO2} and/or \ce{H2O}. ODAC23 is by far the largest dataset of MOF adsorption calculations at the DFT level of accuracy currently available. In addition to probing properties of adsorbed molecules, the dataset is a rich source of information on structural relaxation of MOFs, which will be useful in many contexts beyond specific applications for DAC. A large number of MOFs with promising properties for DAC are identified directly in ODAC23. We also trained state-of-the-art ML models on this dataset to approximate calculations at the DFT level. This open-source dataset and our initial ML models will provide an important baseline for future efforts to identify MOFs for a wide range of applications, including DAC.

\end{abstract}

\end{@twocolumnfalse}
]

\clearpage
\section{Introduction}

%Credit to Logan for a lot of this text and references!

Annual anthropogenic carbon emissions reached nearly 36 billion tonnes in 2020, and the atmospheric carbon dioxide concentration has increased $\sim$50\% since preindustrial times to approximately 420 ppm. \cite{Cheng2022} Rising \ce{CO2} levels have motivated the development of carbon capture and sequestration (CCS) technologies to combat the effects of emissions on global climate change. \cite{Sood2017}  Direct air capture (DAC) is an emerging technology with the potential for distributed capture and negative emissions.\cite{SanzPrez2016} DAC operates at ambient conditions and avoids impurities that are common for point source capture of \ce{CO2}, but the low concentration of \ce{CO2} requires the movement of large volumes of air and strong adsorption of \ce{CO2}.\cite{Realff2012} Many current DAC absorbents, such as liquid amines and solid alkali hydroxides, strongly bind \ce{CO2} through chemisorption, requiring energy-intensive regeneration of the sorbent. \cite{Tiainen2021, Sholl2016}
Metal-organic frameworks (MOFs) are a promising class of alternative sorbent materials for DAC allowing regeneration at relatively low temperatures. In contrast to sorbents such as alkali hydroxides, MOFs are modular, flexible, and highly tunable, and they possess remarkably high porosities, low densities, and long-range order. \cite{Farha2010} The chemical tunability and long-range order make MOFs worthy of high-throughput computational screening studies.

Computational materials design is a promising strategy for DAC sorbents. \cite{Park2020} Design of efficient DAC processes may require tailoring of materials to the specifics of the air temperature and humidity conditions in a given environment or the temperature/pressure swings that are required to keep energy consumption low. \cite{Kim2022} This is particularly true of DAC processes that seek to leverage air movement and energy content of existing systems such as heating, ventilation, and air conditioning. \cite{Leonzio2022} The consideration of humidity is particularly important, since dehumidifying air requires significant energy input, the presence of \ce{H2O} can result in competitive adsorption even at low relative humidities, and humidity can in some cases cause adsorbent degradation over time. \cite{Boyd2019,Findley2021,Chen2022,You2018} The availability of large datasets of MOFs and other solid sorbent materials can facilitate identification of specific materials or chemical moieties that are well suited for the specific conditions of a given DAC process.\cite{Wilmer2012_structProp,Daglar2020}

High-throughput computational studies and machine learning (ML) techniques are already a common practice in the screening and discovery of MOFs and other reticular materials.\cite{Lin2012,Kim2013,Matito-Martos2014,Yilmaz2015,Tang2018,Yan2019,Lee2021,Findley2021,Aydin2022,Schwalbe-koda2021} There are several large databases of MOF\cite{Chung2014,Chung2019,Wilmer2012,Majumdar2021,Colon2017,Rosen2021} and zeolite structures\cite{Pophale2011} and multiple computational toolkits\cite{Dubbeldam2016,Sharma2023,Simon2016,Thompson2022,Shah2017} and ML models\cite{Simon2015,Pardakhti2020,Yu2021,Yao2021,Bucior2019a,Yang2022,Lee2021} to analyze and predict the adsorption properties of these materials. However, there are several key limitations to the existing body of work. First, because of the computational costs involved, many studies rely on empirical force field (FF) models for predicting adsorption properties. Inaccuracies associated with FFs can lead to both qualitative and quantitative inconsistencies in the prediction of material performance, particularly in the case of open-metal sites (OMS) or defects where covalent bonding or complexation occurs.\cite{Lin2012,Dzubak2012,Cleeton2023,Yazaydın2009,Chen2012,Becker2017,Ce2018} There are several large databases of density functional theory (DFT) calculations for MOF materials,\cite{Rosen2021,Chung2019} but to date these are focused only on the MOF structure and do not include adsorption data. Second, many existing databases and studies of \ce{CO2} adsorption focus only on adsorption of \ce{CO2}, neglecting the possibility of competition with \ce{H2O}.\cite{Chung2016,Boyd2019,Deng2020,Bobbitt2023,Burner2023} Failure to consider competitive adsorption will strongly limit the ability to predict materials for practical DAC processes, where bicomponent \ce{CO2}/\ce{H2O} isotherms are required. Accurately modeling \ce{H2O} adsorption  with classical FFs is challenging due to the complex physical properties of water.\cite{Heindel_2023, Steinmann_2018, Brugnoli_2021, Lopes_2006} Third, many computational databases and studies focus on hypothetical materials,\cite{Wilmer2012,Colon2017,Nandy2022,Park2023} which leads to practical challenges in the synthesis and experimental testing of new predicted materials.
Finally, most datasets are restricted to pristine materials. In reality, MOFs will contain a wide range of defects that may govern their adsorption properties under practical conditions.\cite{Zhou2023,Niu2023} New materials can also be created by inserting defects in MOFs via so-called defect-engineering.\cite{Moslein2022}  Large datasets of high-quality DFT simulations of mixed \ce{CO2} and \ce{H2O} adsorption on realistic pristine and defective MOFs are needed to address these limitations.

ML is also a well-established approach in the discovery of MOFs and other nanoporous materials. ML models have been applied to directly predict adsorption properties and isotherms of MOFs based on their physical and chemical structure.\cite{Bucior2019a,Lee2021,Gurnani2021,Anderson2020,Yang2022,Yao2021,Choudhary2022} Descriptors based on the porosity, chemical constituents, and energy landscape of probe adsorbates in MOFs have been combined with a range of regression and classification models to provide predictions of gas loadings,\cite{Fernandez2013,Li2021,Bucior2019a,Lee2021} Henry's constants,\cite{Yu2021,Choi2022} and temperature-dependent isotherms.\cite{Gurnani2021, Anderson2020} Neural networks have been used to predict MOF properties and perform inverse design tasks to identify MOF materials with high thermal stability\cite{Nandy2021,Nandy2022} and strong or selective \ce{CO2} adsorption.\cite{Park2023,Yang2022,Choudhary2022} ML models have also been trained to provide insight into synthesizability and stability of MOFs and zeolites.\cite{Moghadam2019,Batra2020,Luo2022,Jensen2019,Moliner2019} However, the training data required for many of these properties, such as adsorption isotherms, are generated using classical FFs, which have been shown to exhibit systematic errors.\cite{Cleeton2023} Efforts to train ML models that can directly emulate DFT data for MOFs are more limited.\cite{Yang2022} The ability to use ML models to directly replace FFs in MOFs has the potential to enhance many of the prior efforts.

\begin{figure*}[t!]
    \centering
    \includegraphics[width=0.8\textwidth]{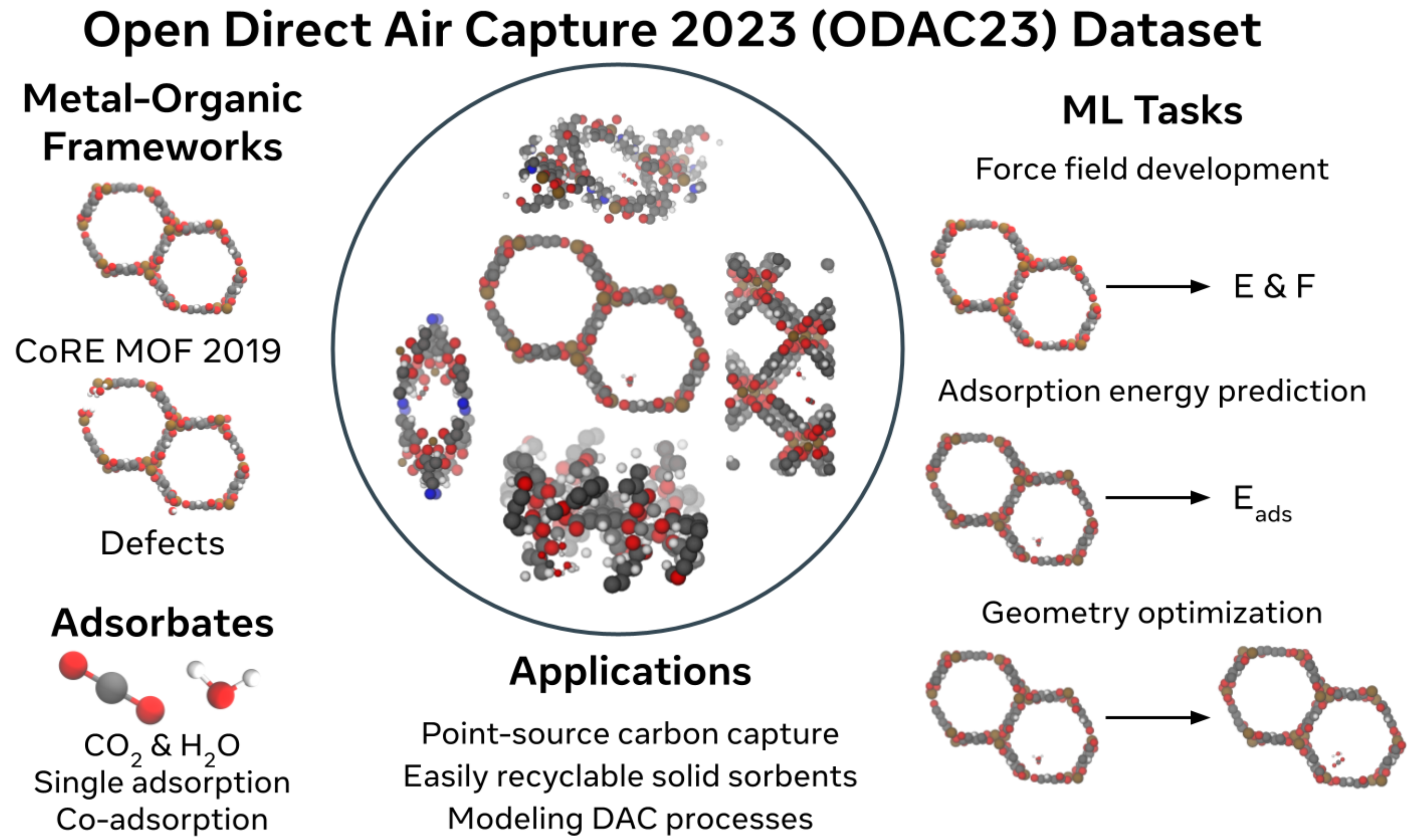}
    \caption{Materials, adsorbates, tasks, and potential applications of the ODAC23 dataset. Images are randomly sampled from the dataset.}
    \label{fig:overall}
\end{figure*}

In this work, we introduce the Open DAC 2023 (ODAC23) dataset to address these challenges. The dataset consists of adsorption energies for \ce{CO2}, \ce{H2O}, and mixtures thereof on $\sim$8K MOFs, amounting to a total of $\sim$176K adsorption energies and $\sim$38M single-point calculations (Fig.~\ref{fig:overall}).~All calculations were performed using DFT with the PBE+D3 exchange correlation functional, ensuring that covalent and electrostatic interactions are treated quantum mechanically and van der Waals interactions are included with well-established empirical accuracy. Approximately 76K adsorption energies involve MOFs that have missing linker defects, providing a route to predicting the role of defects. The dataset is used to train and evaluate state-of-the-art ML models for the prediction of adsorption energies and atomic forces using approaches developed for the Open Catalyst Project.\cite{chanussot2021open} In addition, we include several out-of-domain datasets taken from the extended CoRE MOF database \cite{Nandy2022,Nandy2023} to evaluate the ability of the trained models to generalize to unseen topologies and linker chemistries. We expect that this dataset and the associated infrastructure will accelerate the development of MOF materials for DAC by providing a common dataset that far exceeds the size of any currently available dataset, establishing well-defined standards and benchmarks for development of new ML models, and providing accessible pre-trained ML models that enable routine prediction of mixed \ce{CO2} and \ce{H2O} adsorption on MOFs at an accuracy that approaches DFT.

The ODAC23 dataset is publicly available at the OpenDAC website \footnote{\url{https://open-dac.github.io/}}. All of our trained ML models and training code are available in the OCP repository \footnote{\url{https://github.com/Open-Catalyst-Project/ocp}}.

%\section{Results}
%Nature materials seems to prefer more verbose headings

\section{Scope and Structure of the ODAC23 Dataset}
\label{dataset:mof}

%\subsection{Metal-Organic Frameworks}
%\begin{figure*}
%    \centering
    %\includegraphics[scale=0.35]{figures/Fig_dist.png}
 %   \caption{Number of configurations in each test set of ODAC23 for (a) the S2EF task (single point calculations) and (b) the IS2RE and IS2RS tasks (relaxations). Stacked bar plots indicate the number of configurations by the type of adsorbate present as indicated in the legend. }
    %\label{fig:odac_overview}
%\end{figure*}

%Metal-organic frameworks have been the subject of a wide range of studies related to heterogenous catalysis,\cite{Lee2009} gas storage,\cite{Simon2015,Mason2015} separations,\cite{Li2012,Bao2016,Daglar2020,Tang2021} and drug delivery,\cite{Lawson2021} among others. 
%MOFs are built from metal nodes and organic linkers, and their modularity enables the rational design of materials for targeted applications.\cite{Furukawa2013} 
Enormous numbers of hypothetical MOF structures exist, as illustrated by the hypothetical MOF database (hMOF) of Wilmer \textit{et al.}, which contains 138,000 structures.\cite{Wilmer2012} Several other MOF databases have been developed, including the Topologically Based Crystal Constructor (ToBaCCo) database of 13,512 MOFs with 41 unique topologies developed by Colón \textit{et al.}\cite{Colon2017} Perhaps most importantly, Chung \textit{et al.} developed the Computation-Ready, Experimental (CoRE) MOF database\cite{Chung2014} and its 2019 expansion\cite{Chung2019} from experimentally synthesized structures in the Cambridge Structural Database (CSD).\cite{Groom2016} The CoRE MOF database has been the foundation of many studies and extensions, including assignment of DFT-derived point charges,\cite{Nazarian2015} more thorough cleaning by removal of structures with misbonded or overlapping atoms,\cite{Chen2020} and the QMOF database of DFT-derived properties of many CoRE MOF structures.\cite{Rosen2021}

%Each of the approximately 14,000 structures in the CoRE MOF database has been experimentally synthesized, and each entry contains experimentally reported structural information. Nazarian \textit{et al.} supplemented the CoRE MOF 2014 database by assigning high-quality, DFT-derived point charges for nearly 3,000 structures;\cite{Nazarian2016} such charges facilitate FF-based materials screening when electrostatic interactions must be considered.\cite{Ongari2019} Chen \textit{et al.} performed a thorough cleaning of the CoRE MOF database by identifying structures containing misbonded or overlapping atoms in their input files due to imperfect solvent removal procedures in the original CoRE works.\cite{Chen2020} Some recent MOF databases have utilized CoRE MOF as a starting point, such as the Quantum MOF (QMOF) database developed by Rosen \textit{et al.}, which contains quantum chemical properties such as band gaps and densities of states for most of the CoRE MOFs.\cite{Rosen2021} Although no MOF database is likely to be comprehensive, the CoRE MOF database represents the most expansive reported collection of computation-ready structures of experimentally-synthesized MOFs available to date.

The Open DAC dataset uses the CoRE MOF 2019 work as a starting point. This approach is beneficial because the data are readily available and the origin of each MOF in the database in an experimentally reported synthesis partially addresses concerns surrounding practicality when considering candidate MOFs for experimental testing. The CoRE MOF database has also been shown to be more chemically diverse than larger databases of hypothetical materials, which is beneficial for training transferable and generalizable ML models.\cite{Moosavi2020} The CoRE MOF 2019-ASR database contains 12,020 unique structures with accessible data. We only consider MOFs that contain fewer than 1,000 atoms in the unit cell due to computational cost. MOFs with a pore limiting diameter (PLD) of less than 3.3 Å are excluded because a \ce{CO2} molecule (kinetic diameter of 3.3 Å) may experience kinetic limitations in entering such small pores \cite{Chung2019}. With these limitations, 8,803 MOFs serve as our starting point for DFT relaxation. 

We used the Perdew-Burke-Ernzerhof functional\cite{perdew1996generalized} with a D3 dispersion correction\cite{Grimme2010,Grimme2011} (PBE-D3) for all calculations. The generalized gradient approximation (GGA) approach was chosen over more accurate methods such as hybrid functionals or coupled cluster techniques because of the size and diversity of the dataset.
Nazarian \textit{et al.} showed that several different functionals and dispersion corrections perform  similarly when making structural and partial charge predictions on a chemically diverse set of MOFs.\cite{Nazarian2015} 
%The Hubbard \emph{U} correction has been shown to improve the accuracy for open metal sites,\cite{Rosen2020} but we choose to omit this correction since \emph{U} values are empirical and it is difficult to find a consistent set for all metals from the literature.
We did not include a Hubbard \emph{U} correction. Without this correction, PBE systematically overpredicts binding energies on open-metal sites, but \emph{U} values are empirical and are difficult to find for every metal type.\cite{Rosen2020} We ran calculations as spin polarized to capture spin effects associated with open metal sites.
Our work ultimately seeks to push the baseline description of MOFs for DAC from classical FFs to the PBE-D3 level of theory, so we prioritized consistency across a very large number of calculations rather than absolute accuracy.

% The following sections outline the methods by which the Open DAC dataset was constructed. Figure \ref{fig:flowchart} illustrates this process. First, we relaxed all $8,803$ `pristine' structures from CoRE MOF using DFT. In the converged structures, we performed rational defect engineering by systematically introducing missing linker defects over a wide range of defect concentrations
% %; this enables us to explore the effects of unsaturated open metal sites on gas adsorption behavior moving\todo{Anuroop: Check this sentence?}. 
% Note that the term \emph{pristine} in this work refers to structures taken directly from the CoRE MOF database, many of which already contain open metal sites. Next, we placed \ce{CO2} and \ce{H2O} adsorbates with many configurations in all structures, both pristine and defective. Lastly, all MOF + adsorbate configurations were relaxed with DFT as before.

% \begin{figure}
%     \centering
%     \includegraphics[scale=0.18]{figures/flowchart.png}
%     \caption{The flowchart of computing an adsorption energy from a empty MOF structure.}
%     \label{fig:flowchart}
% \end{figure}

The ODAC23 dataset consists of complete relaxation trajectories of \ce{CO2}, \ce{H2O}, and mixtures of \ce{CO2} and \ce{H2O} on MOF structures derived from the CoRE MOF database. We include two classes of MOF frameworks: \textit{pristine} frameworks and \textit{defective} structures with missing linker defects systematically added.\cite{Yu2023} Pristine MOF structures are obtained from the CoRE MOF database without further modification. Approximately 66\% of the pristine MOFs include frameworks with open metal sites. To test generalizability, we also included 114 ``ultrastable'' MOFs from Nandy \textit{et al.} created by fragmenting and recombining linkers and nodes from the original CoRE MOF database.\cite{Nandy2023}. The final dataset includes a total of \Npristine{}  %5,078 
pristine MOFs and \Ndefective{} defective MOFs with defect concentrations ranging from 1-16\%. The MOFs contain a diverse set of 57 metals, with Zn, Cu, and Cd being the most common, and include a mix of monometallic (89\%), bimetallic (10.7\%), and trimetallic ($<1$\%) frameworks. The abundance of various metals is provided in Fig. \ref{fig:metal_distribution}, and the most common linkers are listed in Table \ref{tab:top_org_linker}. The adsorbates were initially placed using classical FFs and Monte Carlo sampling, with $\sim$2-6 placements per framework. The selection of MOFs and adsorption configurations included in the final set are established by pragmatic constraints and practical considerations. In total, the dataset consists of over 170K converged adsorption energies and nearly 40M single point calculations, corresponding to over 400M core-hours of compute time. Details are provided in the Methods section.

The \gls{ODAC23} dataset has been designed to allow training of ML models to approximate DFT calculations, similar to previous work in heterogeneous catalysis (OC20 and OC22).\cite{chanussot2021open,tran2023OC22} We use the same three task definitions used in the OC20 work. These tasks are briefly summarized below, and we refer the reader to the OC20 paper\cite{chanussot2021open} for more detailed descriptions.

In each task, the input structure is a unit cell periodic in all directions containing a MOF with one or more adsorbates. The ground truth targets of forces, energies, and relaxed structures were all calculated using DFT. For energy targets, we used a non-relaxed adsorption energy:
\begin{equation}\label{eqn:ads_energy_simple}
\tilde{E}_{\text{ads}} = E_{\text{system}} - E_{\text{MOF}}
    - n_{\carbondioxide} E_{\carbondioxide} - n_{\water} E_{\water}
\end{equation}
where $E_{\text{system}}$ is the energy of the MOF and adsorbates, $E_{\text{MOF}}$ is the energy of the relaxed MOF structure without an adsorbate, $n_i$ is the number of adsorbate $i$ and $E_i$ is the energy of adsorbate $i$ in the gas phase.  The tilde on $\tilde{E}_{\text{ads}}$ denotes that $E_{\text{system}}$ is not necessarily a relaxed structure. In specific cases where $E_{\text{system}}$ is relaxed, the tilde is dropped and the adsorption energy is denoted as $E_{\text{ads}}$. More details are provided in the Methods section.

\begin{table*}[t]
    \centering
    \renewcommand{\arraystretch}{1.0}
    \setlength{\tabcolsep}{5pt}
    \renewcommand{\arraystretch}{1.0}
    \setlength{\tabcolsep}{6pt}
    \resizebox{0.9\linewidth}{!}{
    \begin{tabular}{lrrrrr}
        \toprule
        Split & \# Pristine MOFs & \# Defective MOFs & \# Total MOFs & \# Total DFT Relaxations & \# Total DFT Single Points\\
        \midrule
% train & 4,744 & 3,438 & 8,182 & 162,224 & 35,871,295\\
% val & 124 & 74 & 198 & 3,998 & 839,565\\
% test-id & 125 & 97 & 222 & 4,669 & 973,515\\
% test-ood (big) & 85 & 19 & 104 & 1,768 & 381,219\\
% test-ood (linker) & 32 & 0 & 32 & 1,182 & 287,125\\
% test-ood (topology) & 62 & 0 & 62 & 1,612 & 472,256\\
% test-ood (linker \& topology) & 20 & 0 & 20 & 579 & 158,773\\
train & 4,537 & 3,287 & 7,824 & 162,224 & 35,871,295\\
val & 121 & 71 & 192 & 3,998 & 839,565\\
test-id & 120 & 93 & 213 & 4,669 & 973,515\\
test-ood (big) & 66 & 19 & 85 & 1,768 & 381,219\\
test-ood (linker) & 28 & 0 & 28 & 1,182 & 287,125\\
test-ood (topology) & 55 & 0 & 55 & 1,612 & 472,256\\
test-ood (linker \& topology) & 15 & 0 & 15 & 579 & 158,773\\
\midrule
Total & \textbf{4,942} & \textbf{3,470} & \textbf{8,412} & \textbf{176,032} & \textbf{38,983,748} \\
        \bottomrule
    \end{tabular}
    }
    \caption{Overview of ODAC23 dataset organised by dataset split, number of MOF frameworks, and number of DFT calculations.}
    \label{tab:mof_profile}
    \vspace{-10pt}
\end{table*}

%All structures used as inputs to machine learning models contain both the MOF framework and adsorbates. 
The energies of these MOF+adsorbate structures were used to train models for three tasks:

\begin{enumerate}
    \item \textbf{Structure to Total Energy and Forces (S2EF)} takes a structure as input and predicts $\tilde{E}_{\text{ads}}$ of the system as well the force on each atom. This task is analogous to training a force field for all atoms in the system.
    \item \textbf{Initial Structure to Relaxed Energy (IS2RE)} takes an initial guess structure as input and predicts $E_{\text{ads}}$ of its relaxed structure. This task is analogous to predicting an adsorption energy from an initial structure.
    \item \textbf{Initial Structure to Relaxed Structure (IS2RS)} takes an initial guess structure as input and predicts the relaxed position of each atom. This task is analogous to geometry optimization.
\end{enumerate}

\begin{figure*}[ht!]
    \centering
    \includegraphics[width=1.0\linewidth]{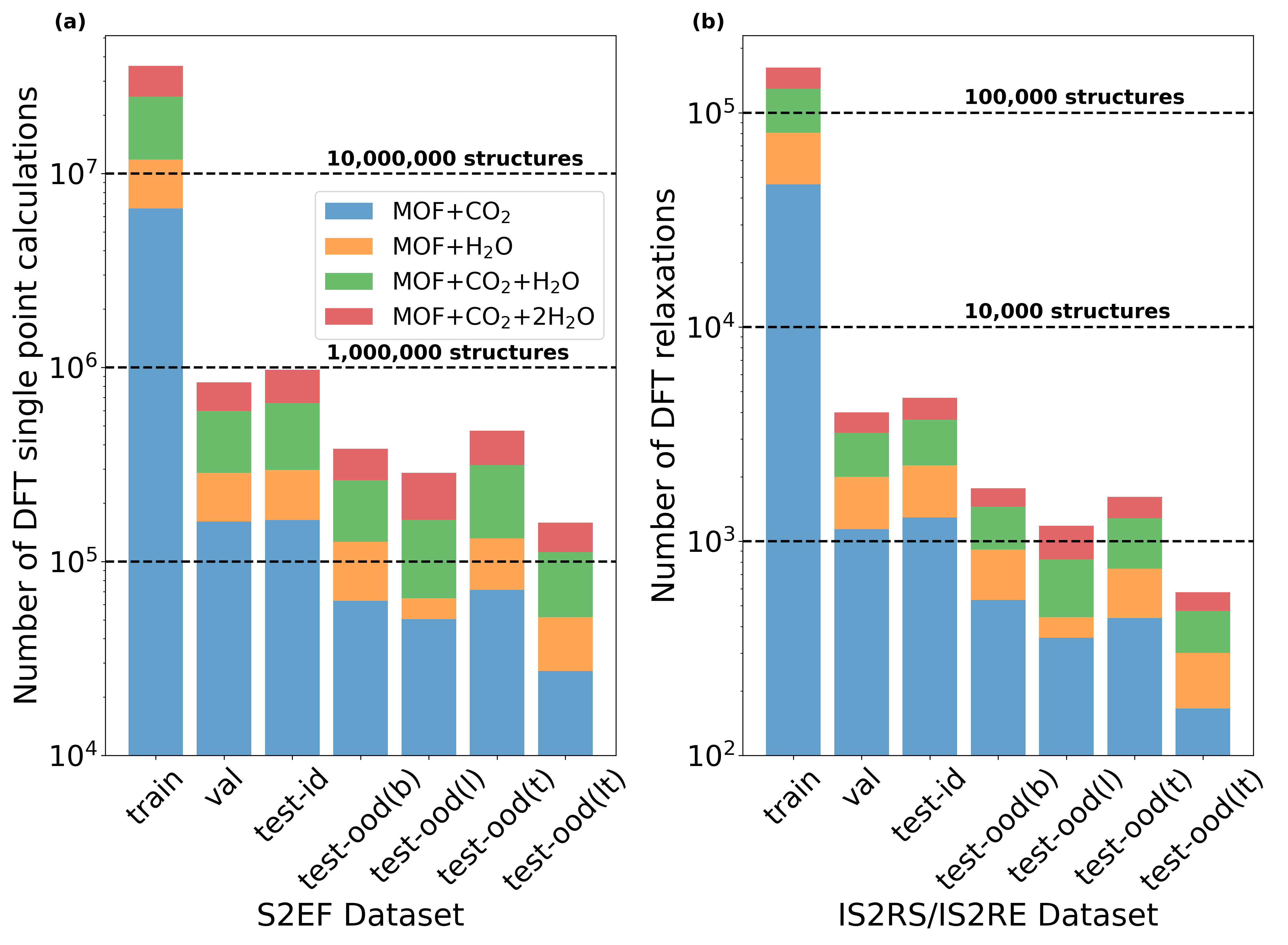}
    \caption{Distribution of the number of MOF+adsorbate DFT calculations for the (a) S2EF and (b) IS2RS/IS2RE tasks on a logarithmic scale. The horizontal lines  emphasize the size of the dataset.
    }
    \label{fig:dist}
\end{figure*}

The S2EF task is the most general, and an S2EF model can be used to complete the IS2RS and IS2RE tasks. The dataset is organized by task and train/test splits. For each task, the data is split into a training set, testing set, and validation set. These in-domain (id) sets are randomly sampled from the full dataset derived from CoRE MOF, but are stratified by MOF framework to ensure that all defective structures are in the same set as the pristine structure from which they are generated. Four out-of-domain (ood) sets are included. The ``big'' ood set corresponds to MOFs from CoRE with over 500 atoms in their unit cell (testing the ability to generalize to larger structures). The ``linker'', ``topology'' ood sets contain linkers and topologies not included in the training data, selected from MOFs in the ultrastable MOF dataset of Nandy \textit{et al.}\cite{Nandy2023} The ``linker and topology'' ood set contains MOFs from the ultrastable MOF dataset that contain both unseen linkers and topologies. The number of MOF structures and DFT calculations in each set is provided in Table \ref{tab:mof_profile}, and a more detailed breakdown based on adsorbate type is provided in \ref{tab:data_profile}. Fig.~\ref{fig:dist} illustrates this detailed distribution across adsorbate types split by task. Further details are described in the Methods section.

\section{Identification of Selective \ce{CO2} Adsorption Sites}
%Finding Promising DAC Sorbents}
\begin{figure*}[ht!]
    \includegraphics[scale=0.35]{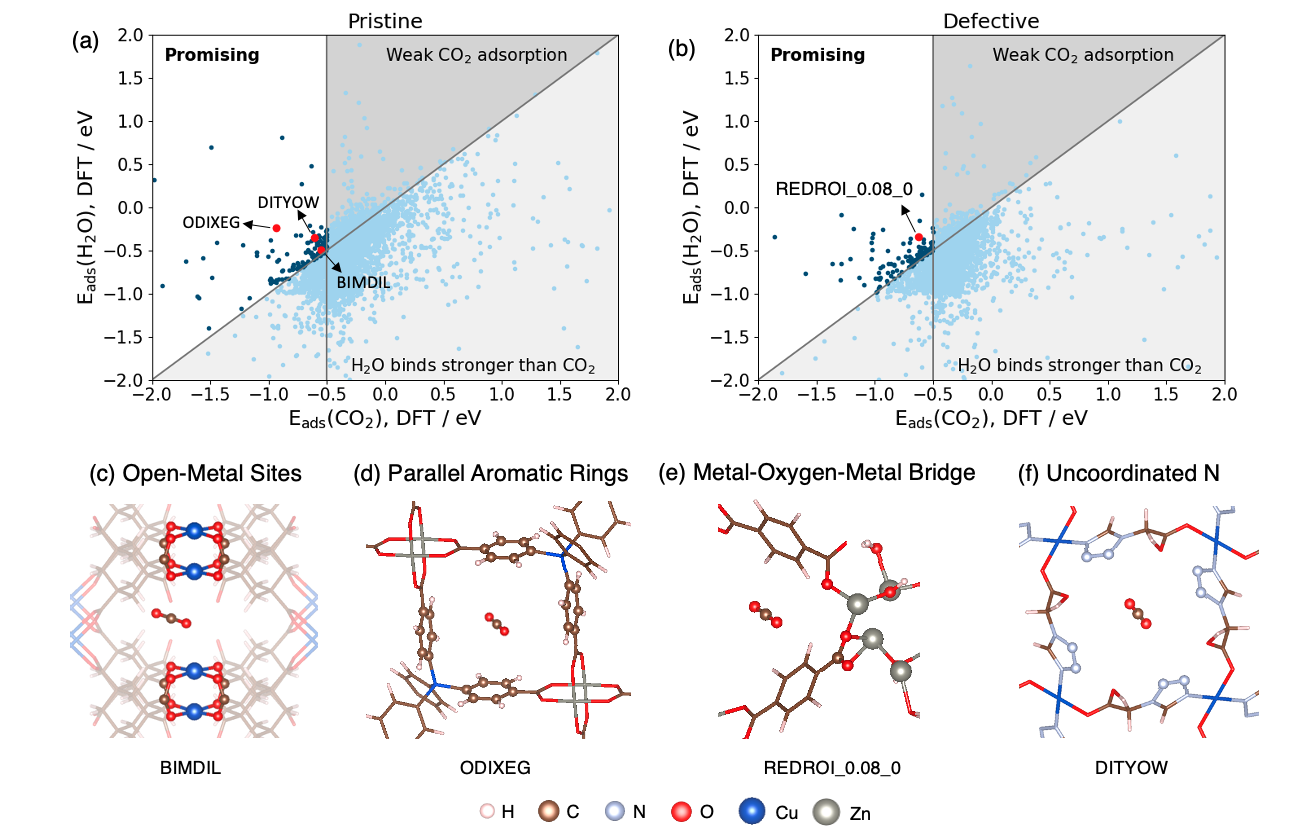}
    \caption{Parity plots showing DFT-calculated ${\carbondioxide}$ and ${\water}$ adsorption energies in (a) pristine and (b) defective MOFs. (c-f) MOF examples with common features of the promising MOFs.}
    \label{fig:Fig_parity}
\end{figure*}

We used our DFT calculations to directly search for MOFs that are potentially interesting for DAC following the criteria suggested by Findley and Sholl\cite{Findley2021} that the adsorption energy of ${\carbondioxide}$ is $<$ $-$0.5 eV (with our sign convention, more negative binding energies correspond to more favorable binding) and that the adsorption energy of ${\carbondioxide}$ needs to be more favorable than that for ${\water}$. Materials not satisfying the first criterion are unlikely to bind sufficient quantities of ${\carbondioxide}$ at the dilute concentrations relevant for DAC, and materials not satisfying the second criterion are likely to adsorb far more water from air than ${\carbondioxide}$. In the following analysis, we compared the lowest adsorption energy of all computed configurations for each MOF + adsorbate case. We neglected cases with $|E_{\text{ads}}/(n_{\carbondioxide}+n_{\water})|>2$ eV because we suspect these cases are unphysical. 

Fig. \ref{fig:Fig_parity}a and b compare the ${\carbondioxide}$ and ${\water}$ adsorption energies in each pristine and defective MOF from our DFT calculations. As expected, most of the MOFs bind water more favorably than ${\carbondioxide}$. However, 135 of the \npristine~pristine MOFs bind ${\carbondioxide}$ strongly and have higher affinity for ${\carbondioxide}$ than for ${\water}$. The top 10 pristine MOFs identified by our DFT calculations with the highest values of $|E_{\text{ads}}(\carbondioxide)-E_{\text{ads}}(\water)|$ are tabulated in Table \ref{tab:promising_pristine}. %\todo{\small{Let's use a notation like $E_{ads}(\carbondioxide)$  unless it's non-standard. Also define this somewhere}}.

Several screenings of the CoRE MOF database for ${\carbondioxide}$ capture in the presence of water have been conducted previously.\cite{Li_Chung_Snurr_2016,Li_Rao_Chung_Li_2017} Here, we compare our promising MOFs with two previous studies where the adsorption energies of ${\carbondioxide}$ and ${\water}$ in CoRE MOFs are available. Findley and Sholl performed a similar screening of CoRE MOFs using FF methods, finding no cases that satisfied the criteria stated above. \cite{Findley2021} The observation that our DFT calculations of analogous quantities identified many interesting materials suggests that the generic FFs used previously are insufficiently accurate. Kancharlapall and Snurr recently screened the CoRE MOF 2019 database with a combination of FF and DFT calculations, using somewhat different selection criteria.\cite{Kancharlapalli_Snurr_2023} 
%Kancharlapall \textit{et al.} selected MOFs with PLD $<$ 6 $\angstrom$,\cite{Li_Chung_Snurr_2016} while we focused only on MOFs with PLD $>$ 3.3 Å, and they limited their search to materials with $<$ 500 atoms per unit cell. 
%All ${\carbondioxide}$ FF-based adsorption energies of the promising MOFs reported by Kancharlapall and Snurr’s work were more positive than $-$0.5 eV, meaning they did not satisfy the criteria we imposed. Kancharlapall and Snurr also used DFT to relax selected configurations in calculations that held the MOF framework fixed. The configuration with the minimum energy was then fully relaxed and treated as the adsorption energy. Our DFT calculations fully relaxed configurations, resulting in greater diversity. 
%17 materials appear in  both our calculations and the list from Kancharlapall \textit{et al.} Our calculations, however, identified adsorption energies for ${\water}$ that were more favorable than for ${\carbondioxide}$ for 10 of these 17 MOFs, meaning that by our criteria these materials are not promising candidates for DAC.
Kancharlapall and Snurr also found that FF-based calculations failed to identify MOFs that satisfy our criteria. They further analyzed a subset of their most promising structures using DFT, with a slightly different workflow than we use for ODAC23. We find that 17 materials identified by Kancharlapall and Snurr also appear in the ODAC23 dataset, though we find that 7 of these materials bind \ce{H2O} more strongly than \ce{CO2} and the remaining 10 MOFs bind \ce{CO2} weakly ($E_{\text{ads}}({\carbondioxide})\geq-0.5$ eV), indicating that they may not be promising for DAC.

In addition to considering the adsorption of single \ce{CO2} and \ce{H2O} molecules, we also used DFT to to probe the co-adsorption of \ce{CO2} and \ce{H2O} in MOFs.  With the resulting co-adsorption energies, we computed the adsorbate-adsorbate interaction energies associated with removing both molecules from the co-adsorbed state, denoted $E_{\text{inter\_mol}}^{\text{1st}}$, for each MOF using equation (\ref{eqn:inter_energy_1st}).  For the 10 MOFs listed in Table \ref{tab:promising_pristine} there are three distinct scenarios for this quantity. In a simple case like ZIDBEV, $E_{\text{inter\_mol}}^{\text{1st}} = 0$ eV is small relative to the single molecule adsorption energies, so co-adsorption can be approximated in a simple way as separate adsorption of the two molecules. For MOFs with negative adsorbate-adsorbate interaction energies like IMAGAG ($E_{\text{inter\_mol}}^{\text{1st}} = -0.64$ eV), co-adsorption of ${\carbondioxide}$ and ${\water}$ is strongly favored relative to adsorption of the individual molecules. Positive adsorbate-adsorbate interaction values such as those seen for IPIDUH ($E_{\text{inter\_mol}}^{\text{1st}} = 1.04$ eV) and TUGTAR ($E_{\text{inter\_mol}}^{\text{1st}} = 0.51$ eV) indicate the co-adsorption is much less favorable than adsorption of isolated molecules. In some cases the first adsorbate-adsorbate interaction energies are strongly nonzero (e.g. KOQLUZ, $E_{\text{inter\_mol}}^{\text{1st}} = -2.31$ eV), suggesting that rearrangement of the MOF structure occurred in the co-adsorbed case that was not observed for the individual adsorbed molecules. 

%we define the interaction energy of the two adsorbed species using equation \ref{eqn:interaction_energy}. 

% \begin{equation}\label{eqn:interaction_energy}
%     E_{int} = E_{ads}({\carbondioxide} + {\water}) - E_{ads}(\carbondioxide) - E_{ads}(\water)
% \end{equation}

% we can define the interaction energy of the second water molecule with the ${\carbondioxide}$+${\water}$ pair using equation \ref{eqn:interaction_energy_2}. 

% \begin{equation}\label{eqn:interaction_energy_2}
% \begin{aligned}
%     E_{int}(2^{nd}\ {\water}) = E_{ads}({\carbondioxide} + 2{\water}) \\
%     - E_{ads}({\carbondioxide} + {\water}) - E_{ads}(\water)
% \end{aligned}
% \end{equation}

 For the ${\carbondioxide}+2{\water}$ configurations, we also computed the second adsorbate-adsorbate interaction energy using equation (\ref{eqn:inter_energy_2nd}). This energy is small or negative for all of the 10 promising MOFs listed in Table \ref{tab:promising_pristine}. One example, LEWZET, shows an extremely negative second adsorbate-adsorbate interaction energy of $-$5.48 eV; this occurs because of significant distortion in the relaxed MOF that occurs due to adsorption of a second water molecule. We note that these effects cannot be explored in existing FF-based searches of MOFs, which assume that the MOF structure is unperturbed by adsorbates. It would be challenging, however, to draw in depth conclusions about selection of MOFs from a limited number of DFT calculations. The complexities associated with the changes in MOF frameworks during co-adsorption and the challenges with sampling the many possible placements of co-adsorbed states both point to the need to be able to derive FFs or ML models that allow rapid assessment of large numbers of states to provide a thorough description of co-adsorption. 

 \begin{figure*}[ht!]
    \includegraphics[scale=0.32]{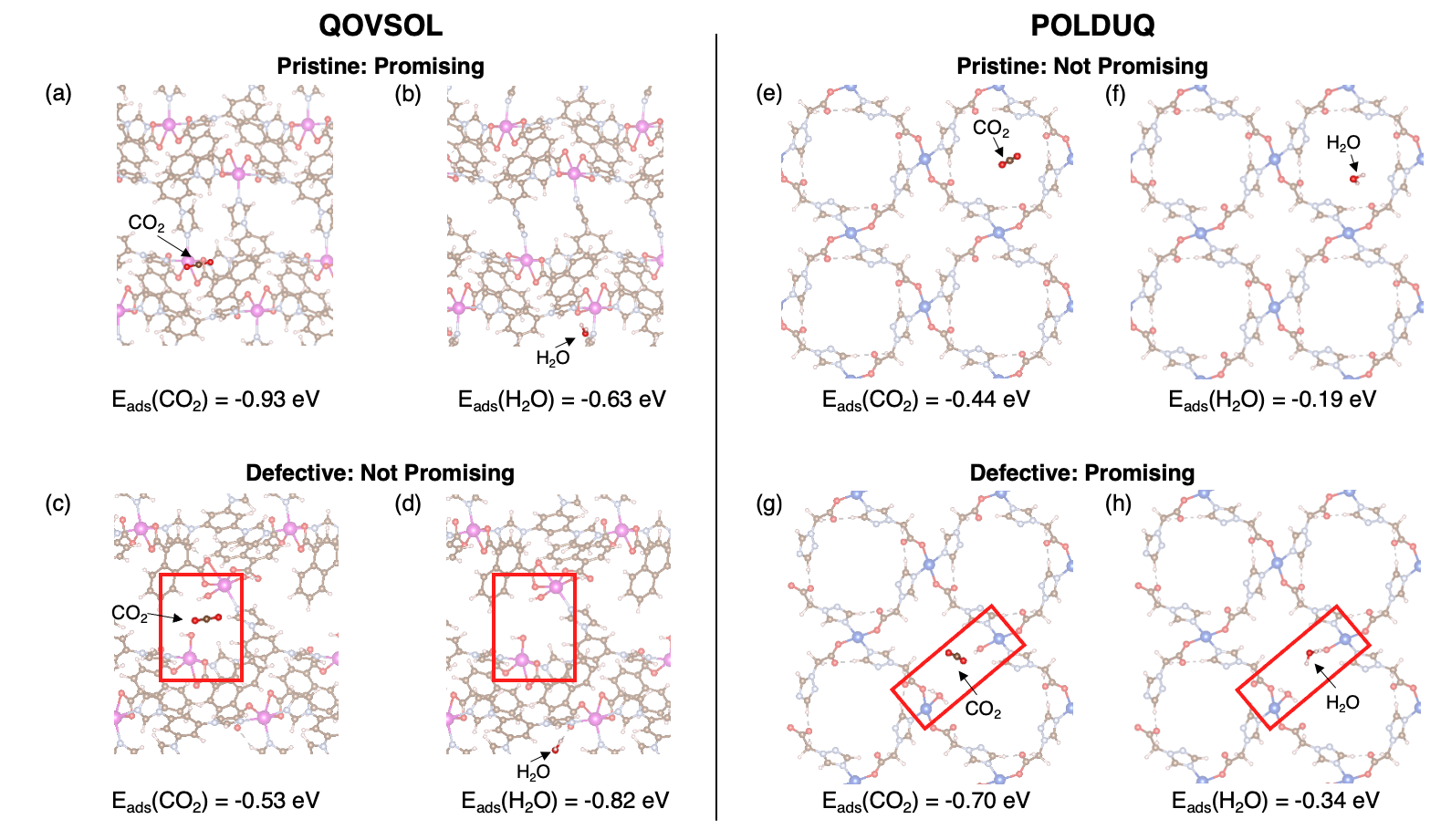}
    \caption{Examples showing different impacts of the defects in MOFs. The defects generated are shown in red squares. Negative impact of defects on DAC (a-d): Defective QOVSOL with a defect concentration of 0.12 shows less favorable \ce{CO2} adsorption (a and c) and stronger \ce{H2O} adsorption (b and d). Positive impact of defects on DAC (e-g): The \ce{H2O} adsorption is slightly more favorable in defective POLDUQ with a defect concentration of 0.06 (f and h), but the \ce{CO2} adsorption in much stronger at the defect site (e and g).}
    \label{fig:Fig_defect_MOF}
\end{figure*}

Our results also include the first large collection of adsorbed molecules in defective MOFs relaxed with DFT. The cell volume of most of the MOFs decreased after introducing defects (Fig.~\ref{fig:Fig_defect_volume}a). From the \ndefective~defective MOFs, we found 107 defective MOFs with \ce{CO2} adsorption energy greater than water (Fig. \ref{fig:Fig_parity}b). The top 10 defective MOFs, ranked in the same way as the pristine materials, are listed in Table \ref{tab:promising_defective}. Defects play an important role in the adsorption of water and \ce{CO2}. For example, pristine TIDLID has adsorption energy of $-$1.10 eV for \ce{CO2} and $-$0.52 eV for \ce{H2O} (Fig. \ref{fig:Fig_defect_volume}b), but defective TIDLID was no longer considered promising because the porous structure collapsed and the PLD was smaller than 3.3 Å (Fig. \ref{fig:Fig_defect_volume}c). 

The defect concentration was not strongly correlated with the difference in adsorption energies associated with the presence of defects (Fig. \ref{fig:Fig_defect_energy}). The average differences of \ce{CO2} adsorption energy were nearly zero for all defect concentrations, and adding defects to MOFs resulted in slightly more favorable water adsorption on average. However, the effect of defects on adsorption energies differs greatly from case to case. In Fig. \ref{fig:Fig_defect_MOF}a to d, defects in QOVSOL resulted in more favorable \ce{H2O} adsorption and less favorable \ce{CO2} adsorption, making it no longer a promising cadidate for DAC. On the other hand, our calculations with defective MOFs show that the defects in some of these materials can create interesting adsorption environments for DAC. We found multiple cases where pristine MOFs would not be selected based on the criteria defined above, but the defective material is a promising candidate. Fig. \ref{fig:Fig_defect_MOF}e to h show one example of POLDUQ. Our observations are broadly consistent with previous experimental and simulation results for \ce{CO2} adsorption in UiO-66,\cite{Hernandez_Impastato_Hossain_Rabideau_Glover_2021,Hossain_Cunningham_Becker_Grabicka_Walton_Rabideau_Glover_2019} and enhanced \ce{CO2} adsorption in Cu-BTC due to water coordinated to OMS. \cite{Yazaydın_Benin_Faheem_Jakubczak_Low_Willis_Snurr_2009} Although defects are capped with water or hydroxyl groups in most cases, it is also possible for defects to create OMSs. The diversity of possibilities illustrates the need for accurate and efficient methods to rapidly explore the many configurations and effects that can exist in defective MOF structures.

It is interesting to ask what motifs or attributes give MOFs adsorption energies that are favorable for DAC. Previous research has suggested several characteristics of good candidates for this application. Boyd \textit{et al.} identified three favorable characteristics: parallel aromatic rings with spacing of approximately 7 Å, metal-oxygen-metal bridges, and open-metal sites.\cite{Boyd2019} The presence of uncoordinated N atoms has also been proposed as a contributing factor to strong \ce{CO2} adsorption.\cite{Liao_CO2_2012,Li_co2_2014} We examined these four characteristics (Fig. \ref{fig:Fig_parity}c-f) in our list of promising MOFs: 224 of the 241 of the promising MOFs can be characterized by at least one of these characteristics, confirming their importance. 
The ODAC23 dataset contains 251 pristine and 267 defective MOFs with an amine functional group. Of these, 7 MOFs (2 pristine and 5 defective) were found to be promising. Structure files of the promising MOFs and the code for promising MOF analysis are available in our open-source repository on GitHub\footnote{\url{https://github.com/Open-Catalyst-Project/odac-data/tree/main/promising\_mof}}.

\begin{table*}[h]
    \centering
    \renewcommand{\arraystretch}{1.0}
    \setlength{\tabcolsep}{5pt}
    \renewcommand{\arraystretch}{1.0}
    \setlength{\tabcolsep}{6pt}
    \resizebox{1\linewidth}{!}{
    %\begin{tabular}{lrrrrrrrrrrrrr}
        %\toprule
        % MOF & E_{ads}(\carbondioxide) & E_{\text{ads}}(\water) & PLD & LCD & Metal & OMS & PAR & M-O-M & Uncoordinated N & \# of citations & \ce{CO2} Loading ($mmol/g^{-1}$) & Ref\\
        %\multirow{2}{*}{MOF} & \multirow{2}{*}{E_{ads}(\carbondioxide)} & \multirow{2}{*}{E_{\text{ads}}(\water)} & \multirow{2}{*}{PLD} & \multirow{2}{*}{LCD} & \multirow{2}{*}{Metal} & \multirow{4}{l}{Binding Sites} & \multicolumn{2}{l}{\ce{CO2} Loading ($mmol/g^{-1}$)}  & \multirow{2}{*}{\# of citations} \\
         
         \begin{tabular}{lrrrrrrcccccr}
         \toprule
\multicolumn{1}{c}{\multirow{2}{*}{MOF}} & \multicolumn{1}{c}{\multirow{2}{*}{$E_\text{ads}$(\carbondioxide)}} & \multicolumn{1}{c}{\multirow{2}{*}{$E_\text{ads}$(\water)}} & \multirow{2}{*}{PLD} & \multirow{2}{*}{LCD} & \multirow{2}{*}{Metal} & \multicolumn{4}{c}{Characteristics} & \multicolumn{2}{c}{Exp. \ce{CO2} Loading (mmol/g)} & \multicolumn{1}{c}{\multirow{2}{*}{\# of Citations}} \\
\cmidrule(l){7-10} \cmidrule(l){11-12}
\multicolumn{1}{c}{}  & \multicolumn{1}{c}{} & \multicolumn{1}{c}{}  &     &   &     & OMS  & PAR  & M-O-M   & Uncoordinated N  &   150 mbar   & \multicolumn{1}{c}{1 bar}   & \multicolumn{1}{c}{}                           \\

          %&  &  & & & & & \multicolumn{1}{c}{OMS} & \multicolumn{1}{c}{PAR} & \multicolumn{1}{c}{M-O-M} & \multicolumn{1}{c}{Uncoordinated N} & \multicolumn{1}{c}{150 mBar} & \multicolumn{1}{c}{1 Bar} & & \\
        \midrule
 ODIXEG & -0.94 & -0.24 & 7.80 & 10.4 & Zn & \cmark & \cmark & &  & & & 56\cite{ODIXEG_syn}\\
 QOVSOL & -0.93 & -0.63 & 3.67 & 6.21 & Cd &  & \cmark & & \cmark & 0.1 (298 K) & 0.2 (298 K)\cite{QOVSOL_isotherm}  & 35\cite{QOVSOL_syn}\\
 QEFNAQ & -0.57 & -0.32 & 4.72 & 6.03 & Cu & \cmark & \cmark &  &  & 0.4 (293 K) & 1.0 (293 K)\cite{QEFNAQ_isotherm}  & 272\cite{QEFNAQ_syn}\\
 FECXES & -0.64 & -0.39 & 6.59 & 10.83 & Cu & \cmark & \cmark &  & & 1.6 (273 K) & 6.3 (273 K)\cite{FECXES_syn} & 56\cite{FECXES_syn}\\
 DITYOW & -0.60 & -0.36 & 4.79 & 4.86 & Cu & \cmark &  &  & \cmark & & & 52\cite{DITYOW_syn} \\
        \bottomrule
    \end{tabular}
}
    \caption{5 pristine MOFs suitable for synthesis on the basis of ODAC23 calculations and manual evaluation of original synthesis reports.}
    \label{tab:synthesis_pristine}
    \vspace{-10pt}
\end{table*}
\begin{table*}[h]
\centering
\renewcommand{\arraystretch}{1.0}
\setlength{\tabcolsep}{5pt}
\renewcommand{\arraystretch}{1.0}
\setlength{\tabcolsep}{6pt}
\resizebox{1\linewidth}{!}{
    %\begin{tabular}{lrrrrrrrrrrrr}
    %    \toprule
    %    MOF & Defect conc. & E_{ads}(\carbondioxide) & E_{\text{ads}}(\water) & PLD & LCD & Metal & OMS & PAR & M-O-M & Uncoordinated N & \# of citations & \ce{CO2} isotherm\\

\begin{tabular}{lrrrrrrccccllr}
\toprule
\multicolumn{1}{c}{\multirow{2}{*}{MOF}} & \multicolumn{1}{c}{\multirow{2}{*}{Defect conc.}} & \multicolumn{1}{c}{\multirow{2}{*}{$E_\text{ads}$(\carbondioxide)}} & \multicolumn{1}{c}{\multirow{2}{*}{$E_\text{ads}$(\water)}} & \multirow{2}{*}{PLD} & \multirow{2}{*}{LCD} & \multirow{2}{*}{Metal} & \multicolumn{4}{c}{Characteristics} & \multicolumn{2}{c}{Exp. \ce{CO2} Loading ($\text{mmol/g}$)} & \multicolumn{1}{c}{\multirow{2}{*}{\# of Citations}} \\
\cmidrule(l){8-11} \cmidrule(l){12-13}
\multicolumn{1}{c}{}  & \multicolumn{1}{c}{} & \multicolumn{1}{c}{}  &  &   &   &     & OMS  & PAR  & M-O-M   & Uncoordinated N  &   150 mbar   & \multicolumn{1}{c}{1 bar}   & \multicolumn{1}{c}{}                           \\

\midrule
POLDUQ & 0.06 & -0.70 & -0.36 & 5.09 & 5.27 & Cu & \cmark &  &  & \cmark & & & 12\cite{POLDUQ_syn}\\
CUGVUW & 0.16 & -1.14 & -0.82 & 3.41 & 5.64 & Cu &  & \cmark &  & \cmark & & &24\cite{CUGVUW_syn}\\
PEPKOL & 0.08 & -0.62 & -0.35 & 3.46 & 3.92 & Ni &  & \cmark &  & & & &444\cite{PEPKOL_syn}\\
SUJNUH & 0.12 & -0.93 & -0.68 & 6.62 & 7.08 & Cu &  & \cmark &  & & 1.3 (195 K) & 2.2 (195 K)\cite{SUJNUH_syn} & 77\cite{SUJNUH_syn}\\
LUYHAP & 0.16 & -0.58 & -0.37 & 8.39 & 12.35 & Cu &  \cmark  & &  &  & & 3.1 (298 K)\cite{LUYHAP_syn}, 2.5 (296 K)\cite{LUYHAP_isotherm}, 5.2 (270 K)\cite{LUYHAP_isotherm}  &158\cite{LUYHAP_syn}\\
   % &  &  &  & &  &  &   & &  &  & &  &\\
   %   &  &  &  & &  &  &   & &  &  & &  &\\

\bottomrule
\end{tabular}
}
\caption{5 defective MOFs suitable for synthesis on the basis of ODAC23 calculations and manual evaluation of original synthesis reports.}
\label{tab:synthesis_defective}
\vspace{-10pt}
\end{table*}

Although the structures in the CoRE MOF set were derived from experiments, it is important to be cautious in concluding that every structure in this dataset is in fact a real material. In developing the CoRE MOF 2019 database, automatic cleaning procedures were applied to experimentally reported crystal structures, including the removal of solvent molecules and the resolution of partial occupancies. Although this procedure was generally effective, there are cases where it was too aggressive. We observed linker removal and wrong partial occupancies in a number of the MOFs listed above. Charge-balancing ions were also removed for MOFs denoted ‘charged’ in Table \ref{tab:promising_defective}. For each MOF listed above, we manually compared the MOF structures retrieved from the CoRE MOF 2019 database and the original publications. From this analysis, we curated a selection of promising MOFs that are completely charge neutral and where the CoRE MOF structure is fully consistent with the original experimental data. On the basis of current DFT data and manual analysis, we expect these to be the most promising MOFs for experimental synthesis and testing. These MOFs are listed in Tables \ref{tab:synthesis_pristine} and \ref{tab:synthesis_defective}. The tables include the number of times the original synthesis report has been cited, since this has been suggested as a proxy for the ease of synthesis/re-use of a material, and the tables indicate which of the four promising MOF characteristics mentioned above appear in each material. The available \ce{CO2} adsorption isotherms from experimental measurements of these MOFs show relatively strong \ce{CO2} adsorption at low partial pressures,\cite{Mahajan_Lahtinen_2022} which is consistent with the implications of our calculations.

\section{Evaluation of the Accuracy of Classical Force Fields}
Our large library of DFT calculations allowed us to further investigate the accuracy of existing classical FFs against our DFT calculations. We focus here on the energy of interaction between adsorbed molecules and MOFs, since this is the key calculation underlying previous high throughput assessments of MOFs for \ce{CO2} adsorption. Specifically, we considered a “standard” FF for adsorption in MOFs  that combines the UFF4MOF,\cite{Rappe1992, Addicoat2014, Coupry2016} TraPPE,\cite{TraPPE} and SPC/E \cite{SPCE_water} FFs for atoms in the MOF, \ce{CO2}, and \ce{H2O}, respectively. Coulombic interactions were defined using DDEC point charges  assigned to MOF atoms from our DFT calculations.\cite{ManzDDEC} Further technical details are provided in the Methods section. 

We computed the interaction energy for 51,478 DFT-relaxed MOF + adsorbate systems using the FF and DFT. This is analogous to the energies in the S2EF task.  Using interaction energies for this comparison rather than adsorption energies is consistent with previous FF-based studies that assume framework rigidity.\cite{You2018, Yu2021_flex, Wilmer2012, Colon2014} The ODAC23 dataset also includes information on MOF deformation associated with the presence of adsorbates, and future work could explore how accurately existing FFs for MOF atoms describe these effects.

\begin{figure*}[ht!]
    \centering
    \includegraphics[width=1.0\linewidth]{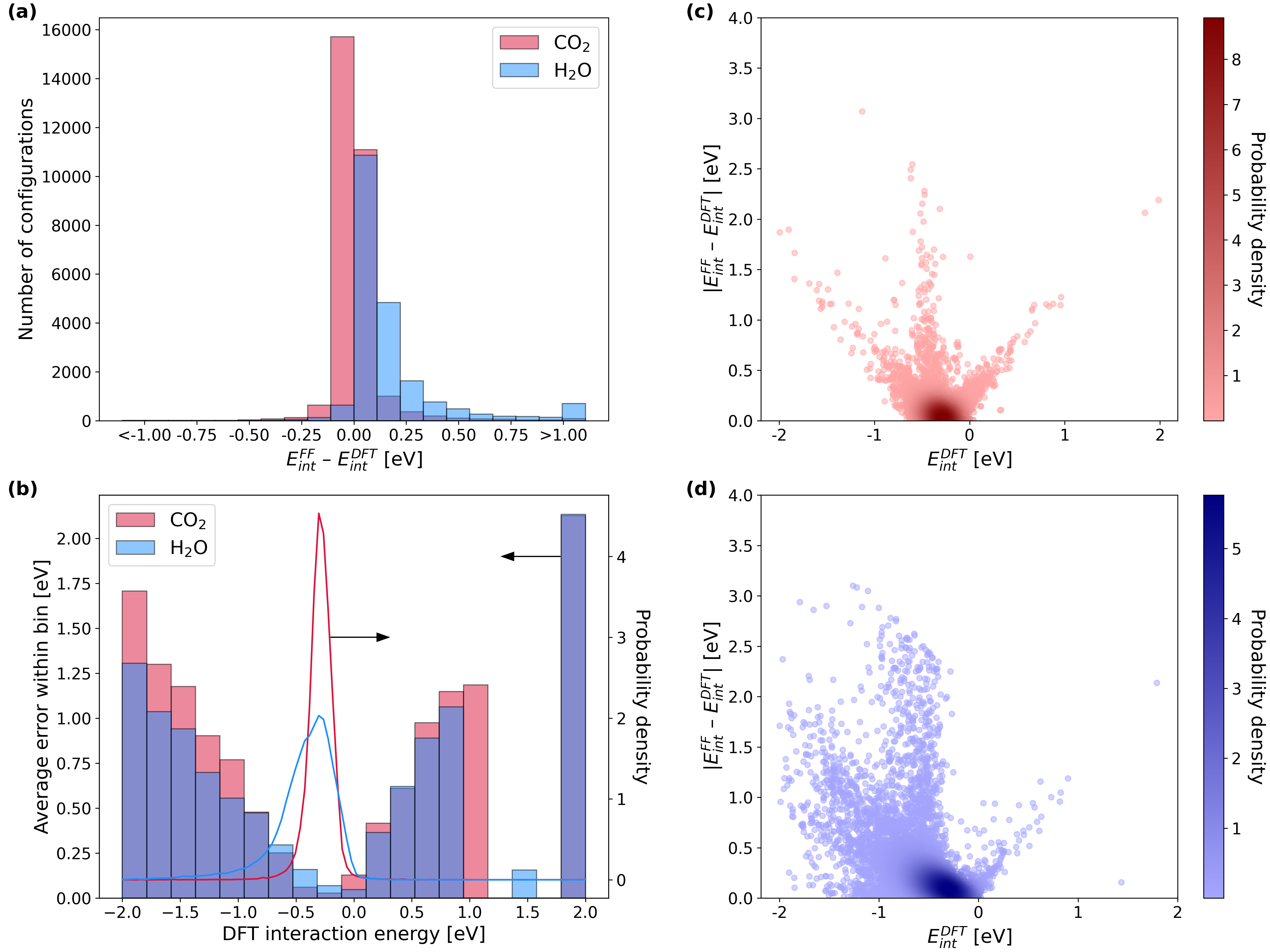}
    \caption{Comparison of adsorbate interaction energies calculated with FFs and  DFT. (a) Histogram of energy differences between FF and DFT for 29,644 \ce{CO2} calculations (red) and 20,892 \ce{H2O} calculations (blue). (b) Binned errors and DFT interaction energy distributions split by adsorbate. (c,d) Absolute difference between FF and DFT plotted  versus DFT interaction energy for \ce{CO2} and \ce{H2O}, respectively. 
    }
    \label{fig:FF}
\end{figure*}

The results of our FF calculations and comparisons to DFT interaction energies are shown in Fig.~\ref{fig:FF}. All structures in this comparison contained only one adsorbate molecule (either \ce{CO2} or \ce{H2O}), and we omit 226 structures with DFT interaction energies outside the range of [-2, 2] eV since we suspect these structures are unphysical. We also omit 716 structures with reasonable DFT energies because their FF predictions also fall ouside of [-2, 2] eV. This is done to avoid heavily skewing the subsequent discussion and is revisited at the end of this analysis. Fig.~\ref{fig:FF}a shows that in many cases the difference between the classical FF and DFT is less than 0.25 eV and that many of the DFT results can be described as physisorption. Van der Waals (vdW) interactions dominate within the physisorption regime of $-$0.5 $\leq$ $E_{\text{int}}^{\text{DFT}}$ $\leq 0$ eV, and if interaction energies are restricted to this range then the mean absolute error (MAE, or simple \textit{error}) between FF and DFT energies is 0.06 %0.061
eV. This indicates that the physics-based FFs we tested are quite well adapted to predict the interaction energy when physisorption is dominant.  

% I probably need to change around this figure formatting to make it more readable, but we can handle that when doing final formatting

The results in Fig.~\ref{fig:FF}b–d provide a less promising view of the classical FF. The error between the FF and DFT calculations scales approximately linearly with the DFT energy outside the physisorption regime, showing that the FF  predicts a physisorption energy even when DFT indicates that chemisorption is occurring. %The FF performs effectively the same as simply returning a constant interaction energy in most cases. 
The minima in these graphs around $-$0.4 eV again indicate that the FF is only capable of accurately predicting physisorption. In the chemisorption regime from –2 to –0.5 eV, the MAEs for \ce{CO2} and \ce{H2O} are 0.29 %0.294
eV and 0.39 eV, respectively. Fig. \ref{fig:FF}b shows the number of points and average error as a function of DFT interaction energy. Although relatively few points outside the physisorption regime exist, the FF interaction energy errors increase drastically with the magnitude of the interaction energy. Many interesting chemistries that are beneficial for DAC occur due to chemisorption (e.g., \ce{CO2} binding more strongly than \ce{H2O}). These cases would be missed by a classical FF that is unable to model chemisorption. There are also many instances in which the FF energy prediction is substantially larger than the DFT-calculated energy. We attribute these to cases involving chemisorption where the adsorbate is close to the framework and therefore returns very large Lennard-Jones energies. That is, the FF exhibits unstable behavior here because very slight changes in geometry cause large spikes in energy predictions. 

An additional takeaway from Fig.~\ref{fig:FF} is that \ce{H2O} is significantly more challenging to model than \ce{CO2}. This is consistent with the fact that physics-based water models are complex and are themselves the subject of a rich body of literature.\cite{Zielkiewicz2005_water} We found that the error in interaction energy calculations within the [$-$2, 2] eV domain involving \ce{H2O} (0.19 %0.188
eV) was more than triple that for \ce{CO2} (0.05 eV). The vast majority of unstable FF calculations involved \ce{H2O} and not \ce{CO2}. Selecting and implementing an appropriate water model is a non-trivial task that further complicates the use of classical FFs for material screening.

Finally, there are a number of cases where the FFs predict very large interaction energies, with the maximum error being 187.2 eV. These cases typically correspond to dissociative adsorption, where the FF is not an appropriate model. Fig. ~\ref{fig:Fig_FF_error_full} presents the binned FF errors as a function of the DFT interaction energy for all configurations with a DFT interaction energy in [-2, 2] eV, irrespective of whether the FF interaction energy falls within this range. Comparison with Fig. \ref{fig:FF}b shows that the 716 cases with reasonable DFT energies but unreasonable FF energies drastically increase the error, and that catastrophic failures (e.g. errors $>$10 eV) begin to dominate when the DFT adsorption energies are stronger than 1 eV. The large errors cause the FF MAE for all structures to be quite large at 0.28 eV. If the MAE is calculated only for cases where the FF interaction energy is in the range of [-2, 2] eV, then the classical FF performs reasonably well with an interaction energy MAE of 0.11 eV across 50,536 calculations. Overall, the results indicate that the FF performs well for physisorption, but fails to capture strong chemical interactions that are likely critical for DAC. 

%If all systems including the 942 originally excluded are included, the MAE jumps to 0.840 eV. % second to last sentence is a poor explanation; suggestions appreciated

%A full scatter plot of errors not limited to the physisorption regime can be found in the SI.

\subsection{Training and Analysis of Machine Learning Models}
% \todo{There is a disconnect between the DFT analysis section, which is entirely focused on adsorption energies, and the ML section, which doesn’t discuss adsorption energies at all (at least not in a plain way). More context needs to be given to the connection between the two sections, and also to why the ML tasks might be useful for things other than considering DAC. (David)}

%\subsubsection{Structure to Energy \& Forces (S2EF)}

We begin by training and benchmarking models for the S2EF task, since it is the most general. We tested six graph neural network (GNN) architectures for this task: SchNet,\cite{schutt2017schnet} DimeNet++,\cite{klicpera2020directional} PaiNN,\cite{schutt2021equivariant} GemNet-OC,\cite{gasteiger2022graph} eSCN,\cite{passaro2023reducing} and EquiformerV2.\cite{liao2023equiformerv2}  We chose models that performed well on the OC20 and OC22 benchmarks since those datasets and tasks are most similar to ours. These models use GNNs containing equivariant or non-equivariant operations to compute energies and forces. 
 All models were trained to minimize the following objective function for forces and energies:
\begin{align}
    \mathcal{L}=\lambda_E\sum_i |\hat{E_i}-E_i| +\lambda_F\sum_{i,j}\frac{1}{3N_i} |\hat{F_{ij}}-F_{ij}|^p
\end{align}
where the loss coefficients $\lambda_E$ and $\lambda_F$ are used to trade-off the force and energy losses. $E_i$ and $\hat{E_i}$ are, respectively, the ground truth and predicted energies of system $i$, and $F_{ij}$ and $\hat{F_{ij}}$ are, respectively, the ground truth and predicted forces for the $j$-th atom in system $i$. The number of atoms in system $i$ is denoted by $N_i$. $p$ is the order of the norm -- SchNet and DimeNet++ used $p=1$, while the other models used $p=2$.

We used the same model sizes as those used for OC20 (Table \ref{tab:model_list}). To prevent overfitting due to the smaller size of the data set, we adjusted the weight decay for each model. We also slightly adjusted the initial learning rates, batch sizes, learning rate schedules, and the loss coefficients $\lambda_E$ and $\lambda_F$. All error metrics are reported for test sets that were not included in the training and optimization process. Additional information can be found in the Methods section.

\begin{figure*}[ht!]
    \centering
    \includegraphics[width=0.8\linewidth]{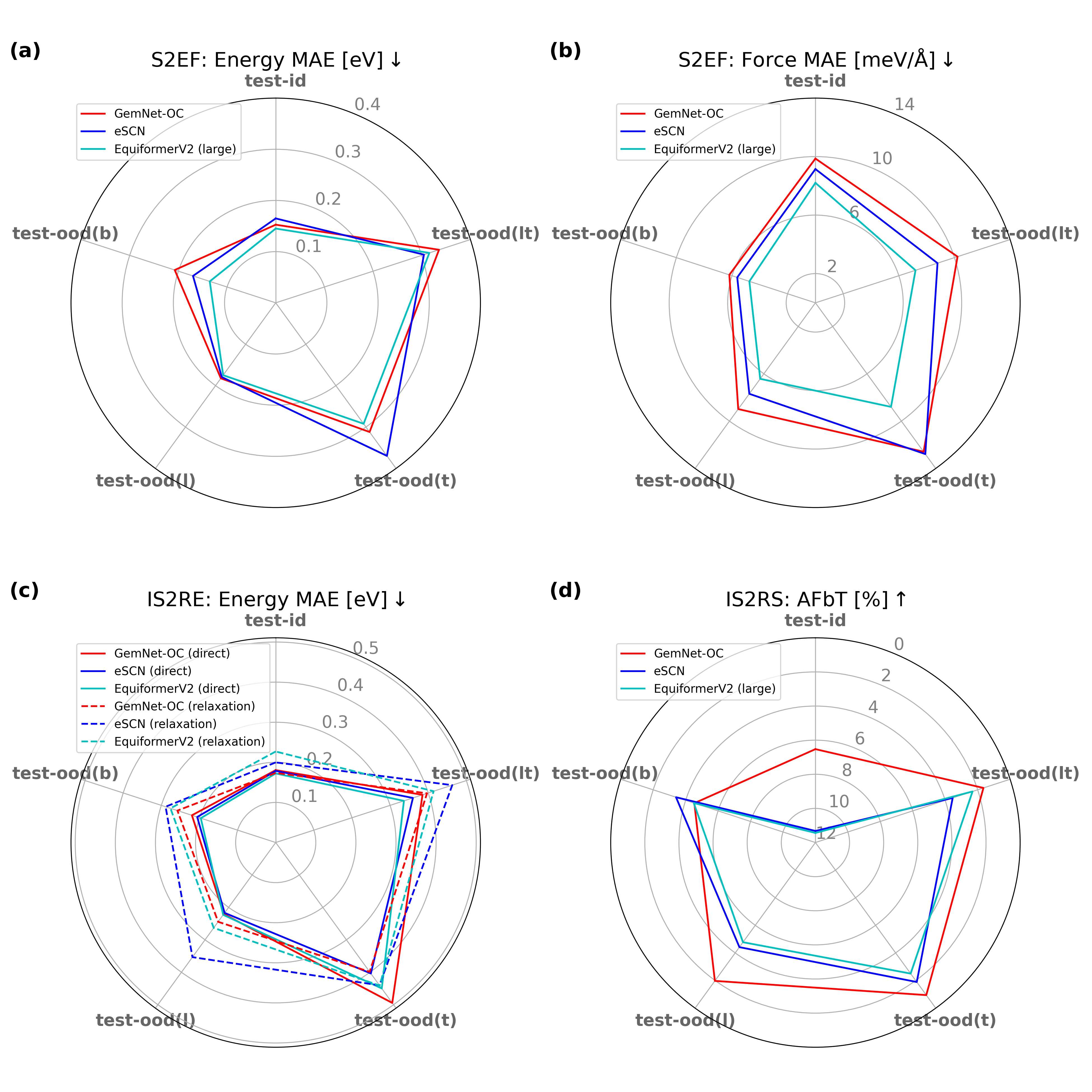}
    \caption{Radar plots for S2EF (a) energy and (b) force  MAEs, (c) IS2RE energy MAEs, and (d) IS2RS AFbT for the top three best models -- GemNet-OC (red), eSCN (blue), and EquiformerV2 (large, except in (c) where the lighter model is shown) (cyan). Dashed lines correspond to the relaxation approach for IS2RE; all other models are direct predictions. Axes correspond to different in- and out-of-domain test sets, and are aligned so that the best result is closest to the origin of the plot in all cases.}
    \label{fig:radar}
\end{figure*}

The results of all ML models on the S2EF task are presented in Table \ref{tab:final_S2EF_subsplit_results}, revealing that GemNet-OC, eSCN, and EquiformerV2 have the best performance. Fig.~\ref{fig:radar} shows a radar plot comparing these models, indicating that%
EquiformerV2 (large) achieved the best results for both forces and energies, with a force MAE of $8.20$ $\meva$~and energy MAE of $0.15$ $\ev$ on the in-domain test set. The eSCN and GemNet-OC models also performed well, with force MAEs of less than $10$ $\meva$~and energy MAEs of under $0.17$ $\ev$. The models' relative performance was consistent with their performance on the OC20 and OC22 datasets, suggesting that improvements in model architecture generalize to various materials datasets.

Next, we consider how the models generalize to out-of-domain test sets. The results in Table \ref{tab:final_S2EF_subsplit_results} and Fig.~\ref{fig:radar} demonstrate that the EquiformerV2 (large) model outperforms the other models on most metrics for all out-of-domain sets. The ML models show only a slight decrease in performance on the \oodb~and \oodl~sets, suggesting that they generalize well to larger graphs or to new linker chemistry. However, the energy predictions for the \oodt~and \oodlt~sets are substantially worse than the \testid~set, although the force errors are similar to the other test sets. This could be due to errors in long-range vdW interactions for unseen topologies, since this is the main contribution that varies with topology. 
%Errors in vdW interactions would have a small effect on the forces, but could accumulate over interactions with many atoms to cause larger energy errors. 
%More research is needed to confirm this hypothesis and improve performance for unseen topologies.  %highlighting the difficulty GNNs have in generalizing to new topologies.

%\paragraph{Effect of open-metal sites \& defects on performance:}

\begin{figure*}[ht!]
    \centering
    \includegraphics[width=0.8\linewidth]{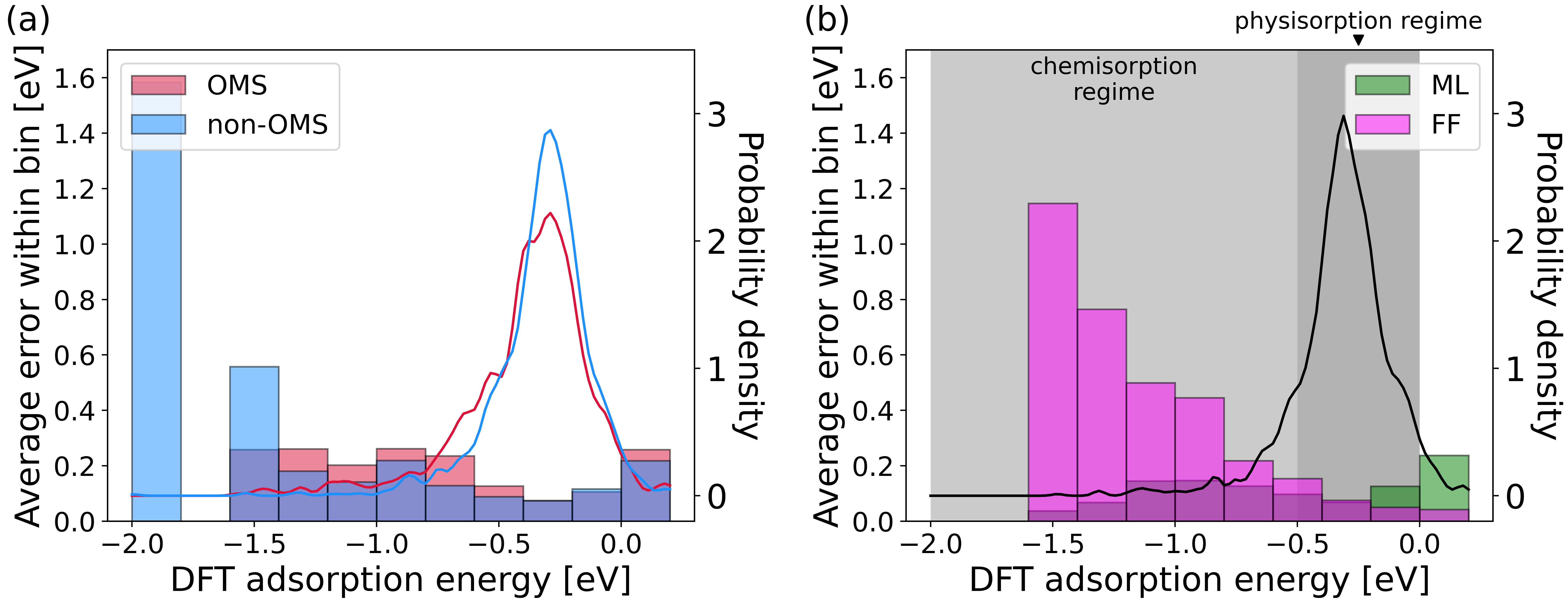}
    \caption{Binned errors and relative density of the number of points (solid lines) as a function of DFT adsorption energy for (a) ML predicted adsorption energies on open metal site (OMS) (red) and non-OMS (blue) and (b) interaction energies predicted by FFs (magenta) and corresponding adsorption energies predicted by ML (green) models. Compared to FFs, ML models are significantly more accurate in the chemisorption regime, and are comparable in the physisorption regime. Positive adsorption energies are omitted from the plot because they are rare and likely unphysical; plots with the full range of adsorption energies are provided in Fig. \ref{fig:binned_plot_si}.}
    \label{fig:ML_vs_FF}
\end{figure*}

We also analyze the performance of the models on the more complex chemical environments of OMSs and defects. OMSs are significant for DAC as they can enable stronger \carbondioxide~adsorption.\cite{Yazaydın2009} Classical FFs are known to be less accurate for MOFs with OMSs as they can cause high polarization in adsorbed molecules.\cite{Yazaydın2009, Kulkarni2016} Tables \ref{tab:final_S2EF_oms_results} and \ref{tab:final_S2EF_defective_results} compare GemNet-OC, eSCN, and EquiformerV2 on different subsets of the \testid~split. Table \ref{tab:final_S2EF_oms_results} shows the performance across pristine MOFs with and without OMSs, and Table \ref{tab:final_S2EF_defective_results} compares the performance of the same models on pristine and defective structures. The ML models have similar force MAEs on the OMS and non-OMS sets, as well as the pristine and defective sets. However, the energy MAEs are lower for MOFs without OMSs or defects. 
This may be due to the stronger and more complex interactions at OMSs, or may be related to the relative abundance of different types of examples within the dataset. Fig. \ref{fig:ML_vs_FF}a analyzes the binned error for MOFs with and without OMSs, indicating that errors are slightly higher for OMS-containing MOFs in the chemisorption regime, suggesting that the ML models perform slightly worse at predicting the more complex chemical interactions at OMS sites.   %{\color{red} We hypothesize that this may be due to the larger variation in adsorption energies for MOFs with defects or OMSs. The standard deviation of adsorption energies for non-OMS and non-defective MOFs is 1.226 and 1.714 eV, while for OMS and defective frameworks it is 1.783 and 1.042 eV respectively.} %Table \ref{tab:final_S2EF_defective_results} compares performance of the same models on pristine and defective structures. Energy MAEs are generally higher on defective MOFs than for pristine MOFs.

% \begin{figure*}[b]
%     \centering
%     \begin{subfigure}[b]{0.48\textwidth}
%         \centering
%         \includegraphics[scale=0.5]{figures/subsets.forces_mae.pdf}
%     \end{subfigure}
%     \hfill
%     \begin{subfigure}[b]{0.48\textwidth}
%         \centering
%         \includegraphics[scale=0.5]{figures/subsets.energy_mae.pdf}
%     \end{subfigure}
%     \caption{Force and energy MAE for the top 3 S2EF models with trained on different amounts of data.}
%     \label{fig:s2ef_train_fractions}
% \end{figure*}

 A direct comparison between classical FFs and ML models is not feasible because the architecture of the FFs makes it challenging to relax framework atoms. However, we can compare the S2EF adsorption energy errors to the interaction energy errors from FFs to gain insight, since both evaluate the ability to describe interactions between frameworks and adsorbates. We did this with 1,391 relaxed single-adsorbate configurations in the test-id set, which is a subset of the 50,536 structures that excludes all systems used in ML model training. For this reason, energy errors reported in this section may vary slightly from those in the evaluation of the accuracy of classical force fields. The energy MAE for EquiformerV2 (large) for these systems was 0.10 eV, while the MAE for the FF interaction energies on the same structures was 0.49 eV. It is clear that, on average, the best ML models outperform the classical FF models, even when only focusing on relaxed single-adsorbate geometries.
% It is evident that, on average, ML models outperform classical force fields significantly.
 %, even on a smaller and simpler subset of relaxed single-adsorbate geometries.

 However, a more detailed analysis reveals that the large FF error occurs due to a small number of large failures. The maximum force field error is 67.66 eV, compared to a maximum error of 1.23 eV for the EquiformerV2 (large) model. If the analysis is restricted to the cases where force fields predict interaction energies in the range of [-2, 2] eV, the average errors are quite comparable, with MAEs of 0.10 eV for both. 
 %In the regime where adsorption energies range from $-$0.5 to 0 eV and physisorption is expected to be dominant, the FFs slightly outperform the ML model, with an MAE of 0.07 eV for the FFs and 0.09 eV for the ML models. 
 In the regime where adsorption energies range from -0.5 to 0 eV and physisorption is expected to be dominant, the FF performance becomes comparable to that of ML, with an MAE of 0.10 eV for the FFs and 0.09 eV for the ML models.
 A detailed analysis is provided in Fig.~\ref{fig:ML_vs_FF}b, which indicates that ML models exhibit consistently lower errors in the chemisorption regime% and are less accurate for repulsive interactions
 , in contrast to FF models, which %perform reasonably well for repulsion but
 fail for chemisorption. Given the importance of chemisorption in selective \ce{CO2} capture at low concentrations, this finding supports the need for ML models for DAC. See Fig.~\ref{fig:binned_plot_si} for errors in the repulsive region.
 
 %On the other hand, the force field models outperform the ML models in the weak bonding regime, with an MAE of 0.067 eV on energies between $-$0.5 - 0 eV, compared to the ML model error of 0.102 eV over the same range. However, the force field model error is much larger for OMS and defective structures 0.268 and 0.782 eV, respectively, while the ML models perform more similarly for these different classes (0.104 and 0.132 eV, respectively). Moreover, the ML models are less prone to catastrophic error. The 90th percentile error for force field interaction energies is 0.247 eV, while the same quantity for ML models is 0.293 eV. These findings confirm that force fields are highly accurate for dispersion interactions, but contradict the conventional wisdom that physical models are better at generalizing and are less prone to catastrophic failure.}

Next, we move to the IS2RE and IS2RS tasks, which evaluate the ability of ML models to directly predict the relaxed adsorption energy (IS2RE) and structure (IS2RS) from an initial guess of framework and adsorbate positions. The IS2RE task only predicts energy and is evaluated with the energy MAE (similar to S2EF) and the ``energy within threshold'' (EwT) which evaluates the fraction of predictions within 0.02 eV of the DFT energy. The IS2RE task can be solved by training ML models to directly predict the relaxed adsorption energy from the initial structure (the \textbf{direct} method), or by running a structure relaxation with an S2EF model (the \textbf{relaxation} method). In the case of the relaxation approach, the task is identical to IS2RS, where the energy of the final structure is used as the IS2RE prediction. However, the metrics used to evaluate the IS2RS task are significantly different, since the goal is to compare structures. The metrics used are the average distance within threshold (ADwT), force below threshold (FbT), and average force below threshold (AFbT), with details provided in the Methods.
Evaluating the IS2RS models is quite expensive since it requires performing a DFT single-point for each of the predicted relaxed structures. Therefore, we only evaluated the best 4 models (GemNet-OC, eSCN, EquiformerV2, and EquiformerV2 (large)) and only computed DFT single-point energies on 500 randomly selected structures from each test split.

%%%%
For the IS2RE task, any S2EF model can be used for the indirect approach, so we evaluated all six S2EF models from this work using the model to perform structure relaxations with each model. The resulting structures are also used for the IS2RS task. In addition, we selected the best
three models -- GemNet-OC, SCN, and EquiformerV2 -- and retrained them for the direct approach, with settings identical to the corresponding S2EF models unless otherwise noted. 
%These models were trained to minimize the mean absolute error (MAE) of the relaxed energy. Unless otherwise specified, the training settings were identical to the corresponding S2EF models. For the relaxation approach, we ran structure relaxations with each of the 6 S2EF models using the L-BFGS optimizer for 125 steps or until the magnitude of the predicted forees on each atom was less than $0.05$ $\eva$.

Fig. \ref{fig:radar} and Table \ref{tab:final_IS2RE_subsplit_results} show the results for the IS2RE task on each of the test splits. On the \testid~set, the direct methods obtain an energy MAE around 0.18 $\ev$ and an EwT of over $10\%$. The relaxation approach with older S2EF models like SchNet, DimeNet++, and PaiNN perform worse than direct methods, while newer methods such as GemNet-OC, eSCN, EquiformerV2, and EquiformerV2 (large) are marginally better than direct approaches. %This is contrary to the findings on OC20 where the relaxation-based approaches are substantially better than direct approaches.
Similar to the S2EF task, we find that the performance of the ML models degrades marginally on the \oodb~or \oodl~datasets, while they degrade significantly on the \oodt~and \oodlt~datasets. This is true for both direct and relaxation-based approaches.% A direct comparison to classical force fields is not feasible for these tasks, since framework relaxation requires reliable topology files for all MOF structures. %However, even for the worst case of \oodlt, the energy MAEs are still substantially lower than the MAE of interaction energies from force fields. 

%\subsubsection{Initial Structure to Relaxed Structure (IS2RS)}
%Finally, we evaluate the models for the IS2RS task. The IS2RS task involves predicting the relaxed state given an initial structure. Similar to IS2RE, this can be done by running structural relaxations using S2EF models and choosing the final structure as the predicted relaxed structure. We used the same relaxation settings as the relaxation method for IS2RE. Evaluating these models is quite expensive since it requires performing a DFT single-point for each of the predicted relaxed structures. Therefore, we only evaluate the best 4 models (Gemnet-OC, eSCN, EquiformerV2, and EquiformerV2 (large)) and only computed DFT single-points on 500 randomly selected structures from each test split.

%The metrics used to evaluate the IS2RS task are significantly different, since the goal is to compare structures. The metrics used are average distance within threshold (ADwT), force below threshold (FbT), and average force below threshold (AFbT), with details provided in the Methods.
Fig. \ref{fig:radar} and Table \ref{tab:final_IS2RS_subsplit_results} show the IS2RS results on each test split.~The ADwT results are reasonably high for the \testid~and \oodb~sets but degrade significantly for \oodl~and \oodt~sets.
%Results from OC20 show that non-DFT distance based metrics like ADwT do not correlate highly with the practical DFT metrics.\cite{chanussot2021open}
eowever, the results on the DFT-based metrics (FbT and AFbT) indicate that the models achieve relaxed structures consistent with what would be obtained from DFT $<$ 1\% of the time in all cases (and 0\% in many cases). This inconsistency between ADwT and (A)FbT has also been observed for OC20\cite{chanussot2021open} and indicates that the models need significant improvement to achieve the level of accuracy needed to replace DFT for the prediction of relaxed structures. However, the fact that the models are able to predict the energies of relaxed structures with reasonable accuracy in the IS2RE task is an encouraging sign, since the state of the art for high throughput MOF screening with force field is to assume that the structures are rigid. This assumption becomes particularly questionable in the case of defective MOFs or strong adsorption, indicating the need for models capable of accounting for relaxation effects.

\begin{figure}
    \centering
    \includegraphics[scale=0.65]{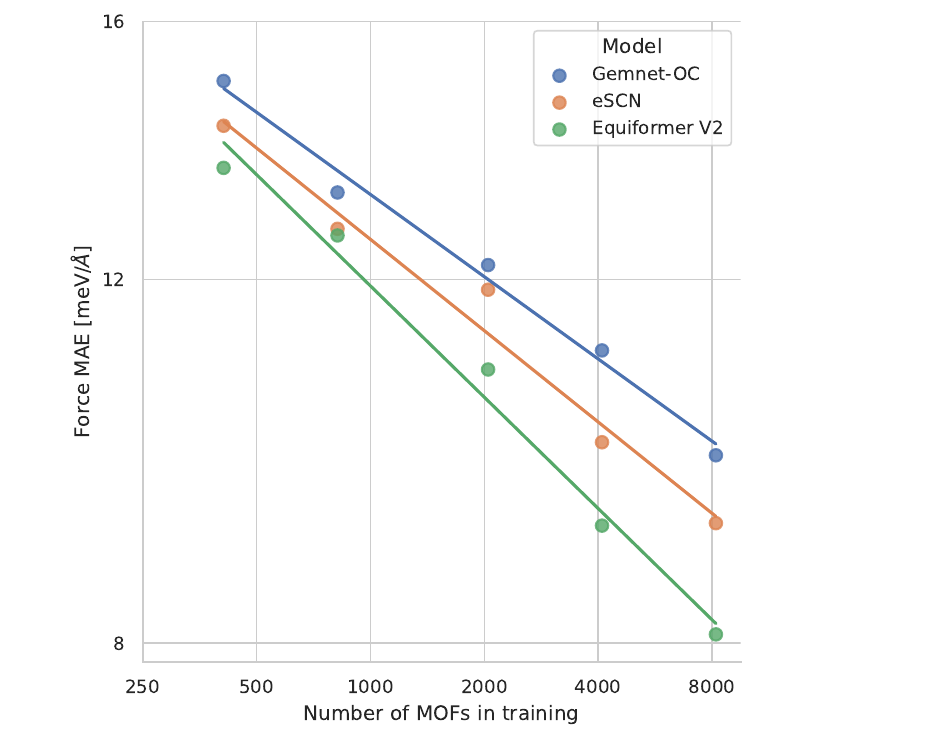}
    \caption{Force MAE on the \testid~set for the top 3 S2EF models when trained on different amounts of training data. The lines show scaling laws obtained by fitting a line between log of the force MAE and log of the number of training MOFs for each model.}
    \label{fig:s2ef_train_fractions}
\end{figure}

%\paragraph{Scaling laws:}
It is clear that the ML models presented here demonstrate significant promise compared to the standard classical FF models. However, there are also obvious deficiencies. One advantage of ML models is that they tend to improve with more data. In particular, scaling laws for deep learning models relate model performance to a parameter like the number of model parameters or size of the training dataset. Scaling laws have helped to choose the optimal model and training parameters in several domains.\cite{hestness2017deep,kaplan2020scaling,hoffmann2022training} Fig.~\ref{fig:s2ef_train_fractions} shows the scaling laws for the ODAC23 dataset size, comparing the force MAEs of different models as a function of the number of MOFs in the training data. Consistent with previous work in other domains, we observe a power-law relationship between force MAE and the number of MOFs. This implies that we can continue to improve the performance of these models by including more training data. It is also interesting to note that equivariant models like EquiformerV2 and eSCN have better scaling properties than GemNet-OC, matching the findings of Batzner \textit{et al.}\cite{batzner20223} This indicates that the use of more sophisticated model architectures is a promising route forward. 

Based on these scaling laws, a much larger number of MOFs would be required to achieve force MAEs of 3 \meva~(approaching the numerical error of DFT). An alternative strategy common in deep learning is to leverage similar datasets. This has proven useful in the Open Catalyst Project models \cite{tran2020methods}, and we plan to explore this approach in future work.
%but transfer learning from OC20 data did not yield any improvements compared to training only on ODAC23. This could be due to differences between the types of materials in the two datasets or due to differences in DFT settings used to create them. 
Another possible strategy is to develop model architectures that are tailored for the DAC application. In particular, the strong performance of FFs in the weak-binding regime suggests that incorporating information on vdW interactions into the model \cite{shuaibi2020enabling} or $\Delta$-ML\cite{ramakrishnan2015big} models may be promising strategies.
Ultimately, we expect that improved model architectures, advanced transfer learning, and joint training techniques ~may provide a route to leveraging physical knowledge and other large atomistic datasets to improve performance on ODAC23, although we leave this as future work.\cite{crawshaw2020multitask,tan2018survey}
%-- for example, Tran et al.~\cite{tran2023OC22} used the OC20 dataset to improve performance on OC22 by fine-tuning OC20 models on OC22 dataset, or by jointly training on both OC20 and OC22 datasets. However, our experiments with similar strategies to leverage OC20 data did not yield any improvements compared to training only on ODAC23. This could be due to differences between the types of materials in the two datasets or due to differences in DFT settings used to create them. Nevertheless, we believe that more advanced transfer learning and joint training techniques \cite{crawshaw2020multitask,tan2018survey}~can successfully leverage OC20 and OC22 datasets to improve performance on the ODAC23 dataset, but we leave this for future work.
\section{Impact and Future Outlook}

The results of this study provide the most comprehensive DFT dataset of \ce{CO2} and \ce{H2O} adsorption in MOFs available.~Analysis of the resulting DFT calculations has shown that, contrary to the findings from FF-based studies, there are numerous MOF-based adsorption sites with strong and selective \ce{CO2} adsorption. A direct comparison of the DFT results to classical FFs provides the most comprehensive perspective to date on the accuracy of FFs. The results reveal that the FFs work well in cases where vdW interactions dominate but fail  when stronger bonding is involved. These findings demonstrate that high-throughput screening with methods capable of treating chemisorption and framework distortion will be required to identify MOFs that can strongly and selectively bind \ce{CO2} under humid conditions.

In addition, the work provides a benchmark for state-of-the-art ML models for \ce{CO2} and \ce{H2O} adsorption in MOFs. The results indicate that the best performing GNN models, such as EquiformerV2, are capable of predicting adsorption energies with average errors of $\sim$0.15 - 0.3 eV, and forces with errors of $\sim$5-10 \meva.~Comparison with a classical FF shows that these ML models are more accurate outside the regime of vdW interactions. This, coupled with the importance of strong binding in identifying selective \ce{CO2} adsorption sites, suggests that these ML models have the potential to replace classical FFs as the standard approach in high-throughput MOF screening for DAC and other applications in separations and catalysis.

Moving forward, it will be important to critically evaluate and improve ML models and associated datasets so that they can be applied to other steps in the computational sorbent selection process. For example, grand canonical Monte Carlo simulations are critical for predicting adsoprtion isotherms. The models here are untested for this task since they have not seen configurations with higher molecular loadings. Testing and improving the models will facilitate calculation of full single and multicomponent isotherms with accuracies that approach DFT. This is especially critical for the case of bicomponent \ce{CO2}/\ce{H2O} isotherms that are needed to predict the behavior of MOF materials in DAC process models.~The complex mixture of vdW, hydrogen, and covalent bonding in \ce{H2O} makes it difficult to accurately predict these bicomponent isotherms with existing methods, but the ML models presented here provide a promising foundation for future developments.

%David -- should we mention anything about experiments here?

\section{Methods}

\subsection{\gls{ODAC23} Dataset Generation }\label{dataset}

A workflow diagram with details on the dataset generation workflow is provided in Fig. \ref{fig:flowchart}, and more details are provided in the sub-sections below.

% \begin{itemize}
%     \item Logan
%     \item MOF background, MOF databases, how we selected MOFs
% \end{itemize}

\subsubsection{Structure relaxations}\label{dataset:relaxation}
% \begin{itemize}
%     \item Anuroop
%     \item Explain DFT settings, adsorption energy calculations, etc.
%     \item I (Logan) moved this subsection from later on in this section because I think it flows better, but feel free to undo. I also tried to write a little bit, but Anuroop should revise as needed
%     \item Do we need to include comparison of geometries like in Nazarian et al. (2017)?
%     \item Perhaps include pyramid figure from Sihoon showing downselection here
% \end{itemize}

%DFT is a widely used quantum chemistry technique and has been extensively documented elsewhere. 

%PBE was used both here and commonly throughout the literature because it is non-empirical, accurate for geometry relaxation in periodic materials, and computationally cheap relative to hybrid functionals for periodic systems. For example, B3LYP is about 2 orders of magnitude more time-consuming than PBE. 
%We included dispersion corrections because conventional DFT cannot accurately describe long-range London dispersion interactions; these are especially important in physisorption of molecules in MOFs and in the details of MOF structures.\cite{Grimme2010} 
DFT relaxations used the PBE exchange–correlation functional\cite{perdew1996generalized} with a D3 dispersion correction\cite{Grimme2010} including Becke-Johnson damping and with spin polarization.\cite{Grimme2011} Relaxations were performed with conjugate gradient methods with a step size of 0.01, and Gaussian smearing was used with a width of 0.2 eV. A plane wave cutoff energy of 600 eV to minimize effects of Pulay stress and a precision of $10^{-5}$ eV were used. All simulations were performed in the Vienna Ab Initio Simulation Package (VASP) v5 software with a 1x1x1 k-point grid.\cite{Kresse1996}

%This further emphasizes that no single approach is unequivocally best and that consistency, convenience, and availability are more important than absolute accuracy in performing DFT calculations of this type.

We relaxed all 8,803 CoRE MOF pristine structures using DFT as described above before generating defective structures and placing adsorbate molecules, and a total of \npristine~MOFs converged. DFT convergence failures are due to a variety of issues. For example, Chen and Manz identified several failure modes in CoRE MOF input files beyond overlapping atoms (3.5\% of all screened structures), including isolated atoms (7.8\%), misbonded hydrogens (1.3\%), and over/underbonded carbons (15.3\%).\cite{Chen2020} Examples of VASP convergence issues were large systems that took too long or ran out of memory ($\sim$10\% of screened structures) and numerical errors pertaining to Hamiltonian diagonalization.  We noticed several converged structures with very high initial formation energies ($>$3 eV/atom). All initial inputs of converged structures were thus screened for overlapping atoms resulting from imperfect solvent removal processes and partial occupancies in the CoRE work. We used the published list of effective atomic radii by Chen and Manz for atom typing; a structure failed if any atom pairs were less than half the sum of their respective atomic radii apart.\cite{Chen2020} In total, 161 structures failed and were excluded from further analysis due to overlapping atoms and unphysically large initial formation energies. 
%The final set of pristine MOFs contained \npristine~structures and serves as the basis for both generating defective structures and performing adsorbate placements below. 
%The DFT-converged structures cover a diverse range of 57 metal types, and the most common types are Zn, Cu, Cd, and Co (see figure \ref{fig:metal_distribution}). 
%The term \emph{pristine} refers to structures taken directly from the CoRE MOF database, many of which already contain open metal sites. 

%Adding the adsorbates into MOFs induced distortions and caused the symmetry breaking of the MOF structures\todo{Anuroop: This is only for some cases, right?}. 
% In some cases, including adsorbates induced structural distortions or symmetry breaking of MOF structures. We conducted a second round of relaxations on some of these MOFs. The adsorbate was removed from the relaxed MOF + adsorbate configuration and the empty MOF was relaxed with fixed cell parameters. We successfully found lower energy states of 690 pristine MOFs and 625 defective MOFs with this procedure. These lower energy states were used as the reference energy for all adsorption energy calculations.

\subsubsection{Defective MOF generation}\label{dataset:defects}
We expanded the pristine set of MOFs from CoRE MOF by introducing missing linker defects using the methods introduced recently by Yu \textit{et al.} \cite{Yu2023} This approach requires identification of the linker and nodes in each MOF, a task completed using the
algorithm MOFid developed by Bucior \textit{et al.} \cite{Bucior2019b} Out of \npristine~pristine MOFs that converged in our DFT calculations, we successfully identified the nodes and linkers of 4,780 MOFs. In each MOF we created structures with different defect concentrations from 0.01 to 0.16, where the defect concentration is defined as the number of removed linkers divided by the total number of linkers. For MOFs that have multiple types of linkers, we generated corresponding defective
structures by removing one kind of linker at a time. OMSs were capped using either a water molecule if the removed linker is charge neutral or hydroxyl(s) if the removed linker was charged to create structures that have no overall charge. In total, 16,358 distinct structures were generated and relaxed by DFT, and 6,340 of them converged. We only kept the relaxed structures with PLD $>$ 3.3 Å, and the final set of defective MOFs contained \Ndefective~frameworks. %This is by far the largest collection of defective MOF structures available to date.

\subsubsection{Adsorbate placement}\label{dataset:adsorbate}
In each relaxed MOF (either pristine or defective) structure, we placed an adsorbate(s) using non-bonded pairwise interactions defined by one of the classical FFs by the RASPA 2.0 package\cite{Dubbeldam2016}. FF parameters for framework atoms and adsorbates (\ce{CO2} and \ce{H2O}) were defined by the United Force Field (UFF)\cite{Rappe1992} and TraPPE-United Atom FF\cite{Potoff2001}, respectively. Specifically, we adopted the rigid TIP5P model for \ce{H2O} molecules\cite{Mahoney2000,Rick2004}. The Lorentz-Berthelot mixing rules and a tail correction with a cutoff radius of 14 $\angstrom$ were used to define the Lennard-Jones interactions between MOFs and adsorbates. Coulombic interactions were considered when partial charges of the framework atoms were available by the DDEC method. We collected configurations of [MOF+\ce{CO2}] or [MOF+\ce{H2O}] from every 10,000 Monte Carlo cycles with the same translation, rotation, and reinsertion probabilities.
%Our desired dataset, which would be a subset of this collection, should avoid having duplicated positions of adsorbates but diverse enough data points in terms of non-bonded pairwise interaction energy. 
We took two approaches to ensure that structures do not have duplicated positions and exhibit diversity in structures: (i) energy matching and (ii) random sampling. The energy matching approach notes that different non-bonded interaction energies will correspond to different configurations. Starting from the minimum observed energy, we sampled configurations in 5 kJ/mol intervals until the non-bonded interaction energy reached a threshold ($-$15 kJ/mol and $-$5 kJ/mol for \ce{CO2} and \ce{H2O}, respectively). If the minimum energy was greater than the threshold, we included only the configuration with the minimum energy. No configuration was added for cases where the minimum energy was $>$ 0 kJ/mol. This resulted in having 0-9 adsorbate placements for each MOF structure, leading to a diverse collection of more than 10,000 MOF+adsorbate configurations per adsorbate by the energy matching approach. For random sampling, we randomly chose 2 configurations from the collection of 10,000 cycles and added these configurations to the set selected from energy matching, leading to more than 16,000 MOF+adsorbate configurations per adsorbate by the random sampling approach. Several MOF structures were further excluded from the dataset because their pore size shrunk during the relaxation, making it impossible for RASPA to place an adsorbate in their pores. We manually added 158 converged [MOF+\ce{H2O}] configurations to position water molecules closer to OMSs. This was done for MOF structures where a \ce{CO2} molecule was near OMSs without nearby water or when they were identified as promising but with fewer than 4 \ce{H2O} placements.
%We manually generated another 158 placements for [MOF+\ce{H2O}] to position the water molecule nearer to the OMSs.

For co-adsorption cases, we used a similar strategy to obtain configurations. Co-adsorption studies include the following examples: [1\ce{CO2}+1\ce{H2O}] and [1\ce{CO2}+2\ce{H2O}]. In each study, we inserted all of the participating molecules into each empty MOF structure. Since we are interested in the behavior of \ce{CO2} in the presence of water, we discarded configurations where the distance between the centers-of-mass for any pair of adsorbate molecules was greater than 5 $\angstrom$.~For MOF structures whose primitive cells were too small to place multiple adsorbates in the pores, the primitive cell was repeated to form a bigger supercell. Whether a supercell was used to save the configuration can be found on GitHub.\footnote{\url{https://github.com/Open-Catalyst-Project/odac-data/tree/main/supercell_info.csv}} Both energy matching and random sampling strategies were applied to multi-adsorbate configurations. The energy threshold was set to be $-$5 kJ/mol for all molecule combinations in case of energy matching approach.

After placing adsorbates, we perform DFT structure relaxations on each adsorbate-MOF configuration. We used the same DFT settings as the MOF relaxations but with fixed unit cell parameters. 
% \begin{itemize}
%     \item Sihoon
%     \item Explain single and multiple adsorbate placement -- MC sampling, how the placements were chosen, etc.
% \end{itemize}

\subsubsection{Out-of-Domain MOF selection}\label{dataset:ood}
The ODAC23 dataset contains four out-of-domain (OOD) test sets in addition to the in-domain test set to evaluate the ability of ML models trained on the ODAC23 dataset to new topologies, new linker chemistries, and to larger MOFs. 

The \textbf{\oodbig} or \oodb~test split only contains MOFs with over 500 atoms in their unit cells. Testing on this set allows us to assess how well our models generalize to larger MOFs than those contained in the training set.

The other three OOD test sets were designed to study how our ML models generalize to new chemistries and topologies not present in CoreMOF. To create these splits, we sampled structures from the ``ultrastable MOF database" developed by Nandy \textit{et al.}\cite{Nandy2023}. %This dataset was created by fragmenting the MOFs from the CoRE MOF into their constituent organic and inorganic components, and then recombining them to create $\sim$54,000 hypothetical MOFs. These MOFs were then filtered down to nearly 10,000 ultrastable MOFs which were predicted to be thermally stable at over 719 K.
To create our OOD test sets, we selected the ultrastable MOFs with less than 500 atoms, and contained either novel linkers or topologies not present in the rest of our dataset. This allowed us to create three OOD test sets: the  \textbf{\oodlinker} set contains novel linkers but known topologies, the \textbf{\oodtopology} set contains novel topologies but known linkers, and the \textbf{\oodlinkertopo} set contains both novel linkers and novel topologies. We abbreviate these three sets as \oodl, \oodt, and \oodlt~respectively. We used the MOFid library\cite{Bucior2019b} to identify the organic linkers and topologies.

We believe that the inclusion of these OOD sets, which are biased to a property not related to the DAC application, provides a useful test of the generalizability of our trained ML models.

\subsection{Energy Definitions}\label{dataset:energy}
We defined three energy definitions for analysis of our work. Throughout this section, $E_{\text{A}}^{\text{B}}$ denotes the total energy of a system of interest $A$ calculated by a method $B$. If not specifically noted, $B$ defaults to DFT. Energies are a function of atomic coordinates (\textit{C}) either from DFT relaxation ($r_{\text{C}}^{\text{relax}}$) or from a single-point DFT calculation ($r_{\text{C}}^{\text{single}}$).
%Moreover, there is a case of $A \subset C$
%while $A \neq C$ where we extracted the atomic coordinates of $A$ out %of a structure $C$.

\paragraph{Adsorption energy}
The adsorption energy can be defined as: 
\begin{equation}\label{eqn:ads_energy}
\begin{aligned}
    E_{\text{ads}} =~&E_{\text{system}}(r^{\text{relax}}_{\text{system}}) - E_{\text{MOF}}(r^{\text{relax}}_{\text{MOF}})\\
    & - n_{\carbondioxide} E_{\carbondioxide}(r^{\text{relax}}_{\text{\carbondioxide}}) - n_{\water} E_{\water}(r^{\text{relax}}_{\text{\water}})
\end{aligned}
\end{equation}
where $E_{\text{system}}$ is the DFT energy of the MOF + adsorbate system, $E_{\text{MOF}}$ is the reference DFT energy of the relaxed standalone MOF,
$n_{\carbondioxide}$ and $n_{\water}$ denote the number of \carbondioxide~and \water~molecules in the system respectively, and $E_{\carbondioxide}$ and $E_{\water}$ are the gas phase energies of the corresponding molecules. 

In equation (\ref{eqn:ads_energy}), the structure of the MOF in the system and the standalone MOF are from separate DFT relaxations. 
%The energy of the standalone MOF is the total energy of the structure before adding the adsorbate. 
When a supercell was created during adsorbate placement, the reference energy $E_{\text{MOF}}$ was computed by performing an additional DFT relaxation on the supercell without the adsorbate.

The inclusion of adsorbate molecules during relaxation broke framework symmetry and resulted in lower energy empty MOF configurations in a small number of cases. We conducted a second round of relaxations on these empty MOFs and successfully found lower energy states for 690 pristine and 625 defective MOFs. These lower energy states were used as the reference energy for all adsorption energy calculations. We removed all configurations where the adsorption energy was found to be $< -2$ eV per adsorbate.

We also define $\tilde{E}_{\text{ads}}$ for which we obtained the total energy of the current MOF + adsorbate configuration instead of seeking its relaxed state. This can be expressed as:

\begin{equation}\label{eqn:ads_energy_tilde}
\begin{aligned}
    \tilde{E}_{\text{ads}} =~&E_{\text{system}}(r^{\text{single}}_{\text{system}}) - E_{\text{MOF}}(r^{\text{relax}}_{\text{MOF}})\\
    & - n_{\carbondioxide} E_{\carbondioxide}(r^{\text{relax}}_{\text{\carbondioxide}}) - n_{\water} E_{\water}(r^{\text{relax}}_{\text{\water}})
\end{aligned}
\end{equation}

\noindent$\tilde{E}_{\text{ads}}$ indicates how far the current state of a MOF + adsorbate system is from its reference state and is used as one of the main targets in our ML studies. In the case that $\tilde{E}_{\text{ads}}$ is computed from the single-point DFT calculation of a relaxed structure (i.e. $r^{\text{single}}_{\text{system}} = r^{\text{relax}}_{\text{system}}$) it is equivalent to $E_{\text{ads}}$. 

% \todo{Nothing said in the text about DFT relaxation of the resulting adsorbate/MOF examples! Need to explicitly say what DFT calculations were performed, what degrees of freedom were relaxed, and what fraction of cases converged. }

% \begin{table*}[ht]
% \begin{center}
% \caption{\# of configurations collected for each adsorption study (I am not sure if we need this actually)}
%     \begin{tabular}{ c | c c c c c c }
%     \toprule
%     \multirow{2}{*}{MOF} & \multirow{2}{*}{\# of structures} & \multicolumn{5}{c}{\# of configurations}\\
%     \cline{3-7}
%      & & $\mathrm{CO_2}$ & $\mathrm{H_2O}$ & $\mathrm{CO_2}$+$\mathrm{H_2O}$ & $\mathrm{CO_2}$+2$\mathrm{H_2O}$ & \textbf{total}\\
%     \midrule
%     pristine & 4,846 & - & - & - & - & \textbf{-}\\
%     defective & 3,723 & 23,191 & 17,363 & 26,029 & 19,147 & \textbf{85,730}\\
%     \bottomrule
%     \end{tabular}
%     \label{tab:configs}
% \end{center}
% \end{table*}

\paragraph{Interaction energy}
The interaction energy is defined as:
\begin{equation}\label{eqn:int_energy}
\begin{aligned}
    E_{\text{int}} =~&E_{\text{system}}(r^{\text{relax}}_{\text{system}}) - E_{\text{MOF}}(r^{\text{relax}}_{\text{system}})\\ 
    & - n_{\carbondioxide} E_{\carbondioxide}(r^{\text{relax}}_{\text{system}}) - n_{\water} E_{\water}(r^{\text{relax}}_{\text{sytem}})
\end{aligned}
\end{equation}
where $E_{\text{int}}$ was calculated either by DFT ($E_{\text{int}}^{\text{DFT}}$) or the classical FF ($E_{\text{int}}^{\text{FF}}$). 
%In each case, all the energies on the right hand side of Eq \ref{eqn:int_energy} were calculated by the corresponding method.

Interaction energies calculations were performed only on the relaxed MOF + adsorbate configurations using single-point DFT. For simplicity, interaction energies were computed only in single adsorption cases, thus $n_{\carbondioxide}+n_{\water}=1$ in equation (\ref{eqn:int_energy}).

%Looking at this quantity, we can understand how attractive (or repulsive) the interactions between adsorbent and adsorbate components are in its final state of the DFT relaxation.

\paragraph{Adsorbate-adsorbate interaction energy}
The adsorbate-adsorbate interaction energy quantifies interactions between adsorbates in co-adsorption cases and is defined as:

\begin{equation}\label{eqn:inter_energy_1st}
\begin{aligned}
    E_{\text{inter\_mol}}^{\text{1st}} =~&E_{\text{ads}}(\carbondioxide+\water)\\
    &- E_{\text{ads}}(\carbondioxide) - E_{\text{ads}}(\water)
\end{aligned}
\end{equation}

\begin{equation}\label{eqn:inter_energy_2nd}
\begin{aligned}
    E_{\text{inter\_mol}}^{\text{2nd}} =~&E_{\text{ads}}(\carbondioxide+2\water)\\
    &- E_{\text{ads}}(\carbondioxide+\water) - E_{\text{ads}}(\water)
\end{aligned}
\end{equation}
where the number of each adsorbate is shown in parentheses. The first adsorbate-adsorbate interaction energy ($E_{\text{inter\_mol}}^{\text{1st}}$) shows the adsorbate-adsorbate interactions between \ce{CO2} and \ce{H2O} and the second adsorbate-adsorbate interaction energy ($E_{\text{inter\_mol}}^{\text{2nd}}$) shows the adsorbate-adsorbate interactions induced by introducing a second \ce{H2O} molecule.

\subsection{Evaluation Metrics}
% \begin{itemize}
%     \item Anuroop/FAIR
%     \item Explain metrics used to evaluate each task.
% \end{itemize}

For all machine learning models we use the same evaluation metrics used for OC20. We briefly describe the metrics used for each task in this section, but refer the reader to the OC20 paper\cite{chanussot2021open} for more details.

\paragraph{Structure to Total Energy and Forces (S2EF)}: The S2EF task is evaluated on the accuracy of force and adsorption energy  predictions through the following metrics. For these metrics $E \equiv \tilde{E}_{ads}$ computed by equation \ref{eqn:ads_energy_tilde}.
\begin{itemize}
    \item Energy MAE: Mean absolute error between the predicted energy and the ground truth energy : 
    \begin{align}
        EMAE = \frac{1}{N} \sum_{i} |\hat{E_i} - E_i|,
    \end{align}
    where $E_i$ and $\hat{E_i}$ are the ground truth and predicted energies of system $i$ and $N$ is the total number of systems.
    \item Force MAE: Mean absolute error between predicted and ground truth DFT forces:
    \begin{align}
        FMAE = \frac{1}{N} \sum_{i} \frac{1}{N_i} \sum_{j} \| \hat{F_{ij}} - F_{ij} \|_{1},
    \end{align}
    where $F_{ij}$ and $\hat{F_{ij}}$ are the predicted and ground truth forces on the $j$-th atom of system $i$ and $N_i$ is the number of atoms in system $i$.
    \item Force Cos: Cosine similarity between the predicted and ground truth forces.
    \item Energy and forces within threshold (EFwT): The fraction of energies and forces that are respectively within $0.02$ $\ev$ and $0.03$ $\eva$~of the ground truth DFT values.
\end{itemize}

\paragraph{Initial Structure to Relaxed Energy (IS2RE)}: The IS2RE task is evaluated on the accuracy of relaxed energy predictions using the following metrics. For these metrics $E \equiv E_{ads}$ computed by equation \ref{eqn:ads_energy}.
\begin{itemize}
    \item Energy MAE: Mean absolute error between predicted energy and the ground truth DFT energy of the relaxed state.
    \item Energy within Threshold (EwT): The fraction of energies within $0.02$ $\ev$ of the DFT relaxed energy.
\end{itemize}

\paragraph{Initial Structure to Relaxed Structure (IS2RS)}: The IS2RS task is evaluated on whether the predicted relaxed structure is close to a local minimum in the energy landscape using the following metrics. 
\begin{itemize}
    \item Average Distance within Threshold (ADwT): Distance within Threshold (DwT) is the percentage of structures with an atom position MAE below a threshold $\beta$. ADwT averages DwT across thresholds ranging from $\beta_0 = 0.01$ $\angstrom$ to $\beta_1 = 0.5$ $\angstrom$ in increments of $0.001$ $\angstrom$.
    \item Force below Threshold (FbT): Percentage of relaxed structures with maximum DFT calculated per-atom force magnitudes below a threshold of $\alpha = 50$ $\meva$. This is only computed for structures that satisfy the DwT criterion with $\beta = 0.5$ $\angstrom$. 
    \item Average Force below Threshold (AFbT): FbT averaged over a range of thresholds: $\alpha_0 = 10$ $\meva$ to $\alpha_1 = 400$ $\meva$ in increments of 1 $\meva$.
\end{itemize}

As the systems in ODAC23 do not contain any fixed atoms, per-atom metrics like Force MAE, ADwT, FbT and AFbT are computed over all atoms. Note that a new single point DFT calculation is required to evaluate FbT and AFbT on a given data point.

\subsection{Classical Force Fields}
%The high computational cost of ab initio methods necessitates approximate parameterizations of the potential energy surface (PES) called force fields (FF) to improve the efficiency of HTS studies. Force fields can be developed to be nearly universal, such as the Universal Force Field (UFF)\cite{Rappe1992} and DREIDING,\cite{Mayo1990} or to be specific to a particular application or chemical environment. The form and choice of FF parameters varies widely, and FF selection can significantly affect the results of HTS studies.\cite{Cleeton2023} Advances in machine learning (ML) have accelerated development of force fields derived from accurate ab initio data; examples include FFs fitted to DFT data using artificial neural networks or genetic algorithms.\cite{Fang2014,Ramprasad2017,Takamoto2022,Mercado2016}

%A necessary step in developing generalizble and transferable models for MOF DAC is to benchmark the performance of existing force fields on our dataset. 
All classical FF calculations in this work used the readily available MOF extension to the ubiquitous UFF force field (UFF4MOF)\cite{Addicoat2014,Coupry2016} in the Large-scale Atomic/Molecular Massively Parallel Simulator (LAMMPS).\cite{Thompson2022} Topology files were generated using LAMMPS Interface.\cite{Boyd2017_lammps_int}  \ce{CO2} and \ce{H2O} molecules were described using the TraPPE\cite{TraPPE} and SPC/E\cite{SPCE_water} models, respectively. The SPC/E model was chosen to avoid challenges related to the geometry of massless sites in newer models such as TIP5P,\cite{Mahoney2000} which was used for adsorbate placement but can be cumbersome to place in high-throughput FF calculations. Electrostatic interactions were described using DDEC framework point charges provided as part of the ODAC23 dataset,\cite{ManzDDEC} and long-range interactions were computed using an Ewald summation with a force tolerance of $10^{-5}$ \text{ kcal/mol/Å}. The cutoff for all pairwise interactions was 12.5 \text{Å}. Periodic boundary conditions were applied in all calculations, and tail corrections were not applied. Code for FF calculations is available in our open-source repository on GitHub\footnote{\url{https://github.com/Open-Catalyst-Project/odac-data/tree/main/force\_field}}.

% This paragraph removed for space
%UFF4MOF adds 92 atom types for metals and coordination environments not included in the original UFF work and more than doubles the number of CoRE MOF 2014 geometries that can be described using UFF parameters to 4,892 of approximately 5,000. UFF4MOF follows the UFF convention that non-bonded parameters are identical for each atom identity, so UFF alone is sufficient for describing adsorbate interactions with framework atoms, though UFF4MOF is needed to relax framework structures. UFF and UFF4MOF are advantageous in their simplicity and generalizability; their functional forms are based solely on elemental identity, hybridization, and connectivity.\cite{Rappe1992} 

% Unlike classical FFs, recent approaches have used ML methods to develop accurate black-box models for both property prediction and structural minimization. One such tool is the Materials Graph Network (MEGNet) architecture developed by Chen et al. based on graph neural networks.\cite{Chen2019_meg}  We used a MEGNet formation energy prediction model that was pre-trained on the Materials Project 2019\cite{Jain2013} database of crystals. A recent update to MEGNet dubbed M3GNet accounts for three-body interactions and includes forces and can therefore perform structural relaxation.\cite{Chen2022_m3g}

% We benchmarked both models on each of the three Open DAC tasks, and the results are shown below. The key takeaways are...

\subsection{ML Models}
% \begin{itemize}
%     \item Anuroop/FAIR
%     \item Models: SchNet, Dimenet++, Gemnet-OC, PaiNN, eSCN, EquiformerV2
% \end{itemize}

Various ML FF models have been proposed for molecular and material tasks over  the last few years \cite{schutt2017schnet,shuaibi2021rotation,sriram2022towards,gasteiger2021gemnet,gasteiger2022graph,schutt2021equivariant,passaro2023reducing,liao2023equiformerv2}. Here, we benchmark a subset of the state-of-the-art models on our tasks. All of our models were implemented using PyTorch \cite{paszke2019pytorch} and the code is available in our open-source repository on GitHub\footnote{https://github.com/Open-Catalyst-Project/ocp}.

%Graph Neural Networks (GNNs) have recently emerged as the dominant method for machine-learning force fields for atomic systems.~Graphs are a natural way to represent atomic systems with each atom represented by a node in the graph, and pairs of atoms that are within a pre-specified cutoff distance are connected by by edges.~This definition can be naturally extended to account for periodic boundary conditions.~Unlike previous approaches that rely on hand-crafted features\cite{}, GNNs learn atomic representations through multiple message passing steps.

%We benchmarked several models that have performed well on the OC20 and OC22 datasets.~
For S2EF, we trained SchNet\cite{schutt2017schnet}, DimeNet++\cite{klicpera2020directional}, GemNet-OC\cite{gasteiger2021gemnet}, PaiNN\cite{shuaibi2021rotation}, eSCN\cite{passaro2023reducing}, and EquiformerV2\cite{liao2023equiformerv2} models. We trained 2 versions of the EquiformerV2 model -- a small 31M parameter model and a large 153M parameter model. The list of models used is summarized in Table \ref{tab:model_list}.Edges were computed on-the-fly using a nearest-neighbor search with a cutoff of 8 $\angstrom$ and a maximum of 50 neighbors for SchNet, DimeNet++ and PaiNN, and a maximum of 20 neighbors for eSCN and EquiformerV2. GemNet-OC uses different cutoffs for different types of interaction triplets and quadruplets.  These S2EF models can then be used to run machine learning relaxations to solve the IS2RE and IS2RS task. We benchmarked the top performing S2EF models -- GemNet-OC, eSCN and EquiformerV2 -- to run these ML relaxations using the L-BFGS optimizer for 125 steps or until the magnitude of the predicted forces on each atom was less than $0.05$ $\eva$. IS2RE can also be solved by directly predicting the energy from the initial system, which we call \textit{direct IS2RE prediction}. We trained GemNet-OC, eSCN and EquiformerV2 models on the direct IS2RE task.

%The SchNet model introduced continuous-filter convolutional layers to apply message passing directly to continuous 3D positions. DimeNet++ proposed directional message passing, and captures interactions between triplets of atoms by transforming the messages based on angular information. This allows the model to better capture the 3D geometry. GemNet-OC extends this to quadruplet interactions, as well as other improvements like symmetric message passing, scaling factors, and equivariant predictions.~PaiNN is an equivariant model that uses spherical harmonics up to the order of $l=1$. The Equivariant Spherical Channel Network (eSCN) model also uses spherical harmonics to represent atom embeddings. By reducing SO(3) convolutions to equivalent, but more efficient SO(2) convolutions, eSCN is able to use higher order spherical harmonics representations, which helps to better model atomic interactions. EquiformerV2 uses the efficient convolutions for eSCN in an equivariant transformer architecture to obtain the current state-of-the-art results on the OC20 dataset.~


% \begin{suppinfo}

% \end{suppinfo}

\begin{acknowledgement}
The authors acknowledge Larry Zitnick (Meta), Joe Spisak (Meta), and Julius Kusuma (Meta) for helpful discussions about the project, Muhammed Shuaibi (Meta) for feedback on dataset construction, and Kyle Michel (Meta) for his help with the compute infrastructure necessary for running DFT calculations.
\end{acknowledgement}

\paragraph{Notice of Copyright}: This manuscript has been authored by UT-Battelle, LLC, under contract DE-AC05-00OR22725 with the US Department of Energy (DOE). The publisher acknowledges the US government license to provide public access under the DOE Public Access Plan (http://energy.gov/downloads/doe-public-access-plan)

\bibliography{main_arxiv}

\clearpage
\onecolumn
\appendix

\pagenumbering{roman}
\setcounter{page}{1}

{\centering
\section{Supplementary Information}
\Large\bfseries The Open DAC 2023 Dataset and Challenges for Sorbent Discovery in Direct Air Capture
\par
}

\setcounter{figure}{0}
\setcounter{table}{0}
\renewcommand{\thefigure}{S\arabic{figure}}
\renewcommand{\thetable}{S\arabic{table}}

%%% Put all the figures here %%%
\subsection{Supplementary Figures}
\begin{figure}[ht!]
    \centering
    \includegraphics[width=0.5\linewidth]{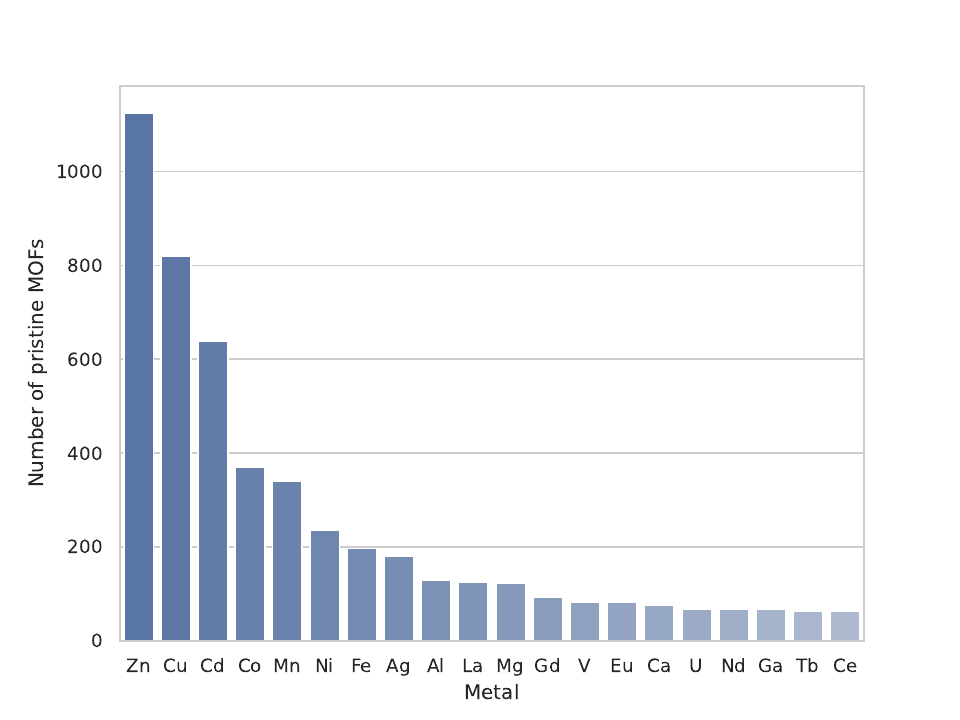}
    \caption{Top 20 metal atoms in pristine MOFs.}
    \label{fig:metal_distribution}
\end{figure}

\begin{figure*}[ht!]
    \includegraphics[width=1\linewidth]{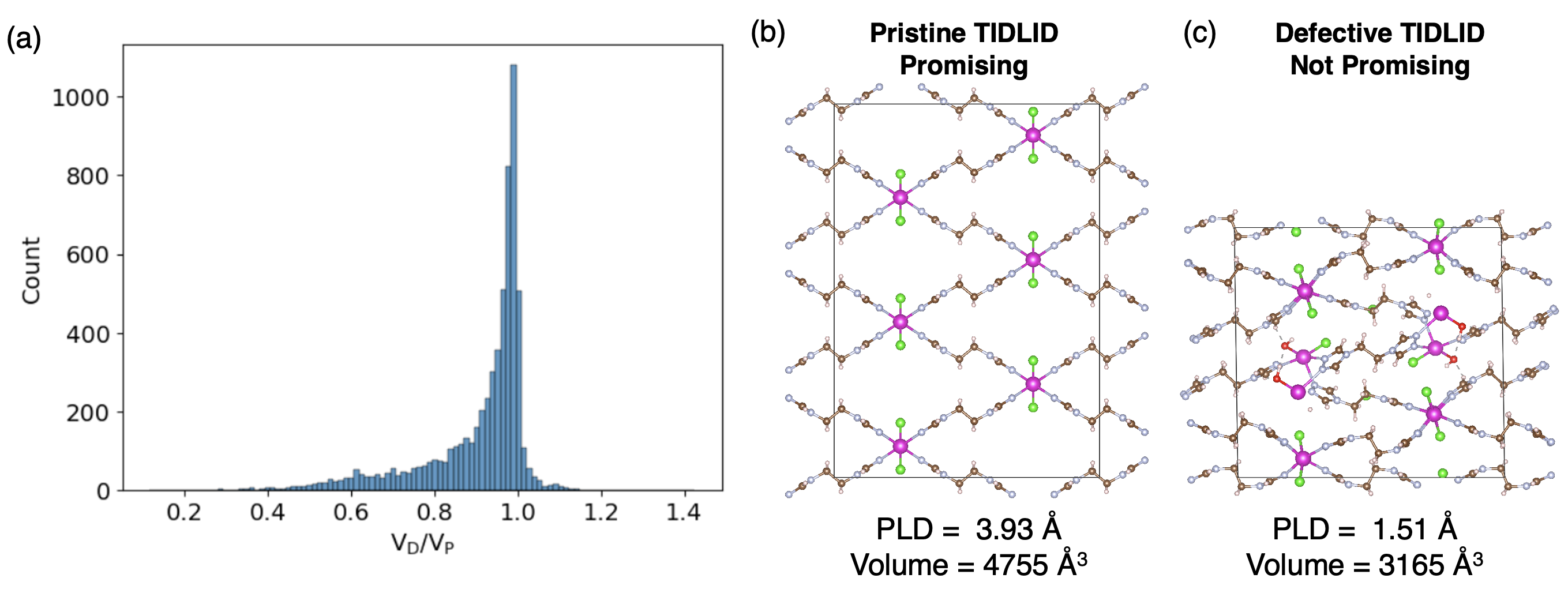}
    \caption{(a) A histogram of the ratio of the defective MOF volume ($V_D$) over the pristine MOF volume ($V_P$), with all structures fully DFT relaxed. The structure of relaxed (b) pristine and (c) defective TIDLID with a defect defect concentration of 0.08 , showing structural collapse on defect formation.}
    \label{fig:Fig_defect_volume}
\end{figure*}

\begin{figure*}[ht!]
    \includegraphics[scale=0.4]{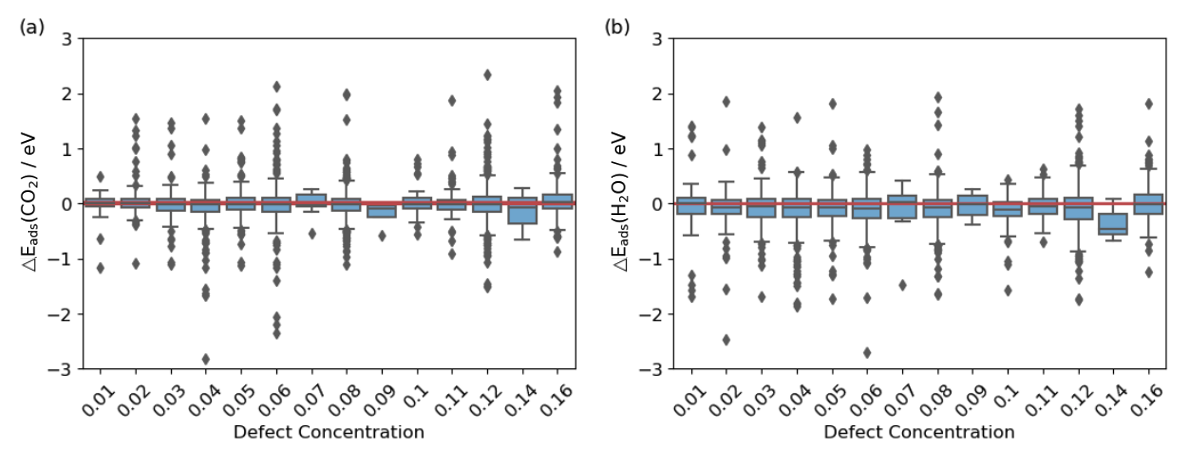}
    \caption{The influence of defect concentration on the adsorption energies of (a) ${\carbondioxide}$ and (b) ${\water}$ in MOFs. $\Delta E = E_{\text{Defective}}-E_{\text{Pristine}}$.}
    \label{fig:Fig_defect_energy}
\end{figure*}

\begin{figure*}[ht!]
    \includegraphics[width=0.8\linewidth]{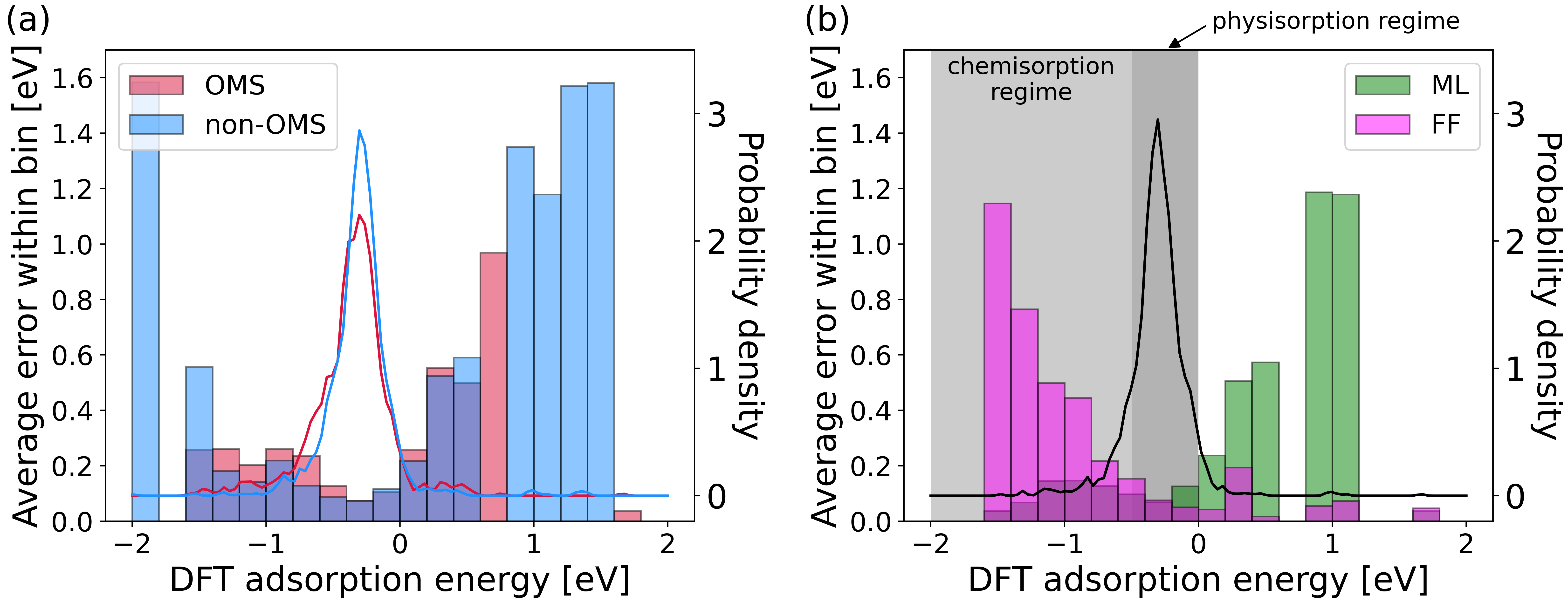}
    \caption{Binned errors and relative density of the number of points (solid lines) as a function of DFT adsorption energy for (a) ML predicted adsorption energies on open metal site (OMS) (red) and non-OMS (blue) and (b) interaction energies predicted by FFs (magenta) and corresponding adsorption energies predicted by ML (green) models. This plot is an extension of Fig. \ref{fig:ML_vs_FF}, displaying DFT adsorption energies within the range of -2 eV to 2 eV.}
    \label{fig:binned_plot_si}
\end{figure*}

\begin{figure*}[ht!]
    \includegraphics[scale=0.4]{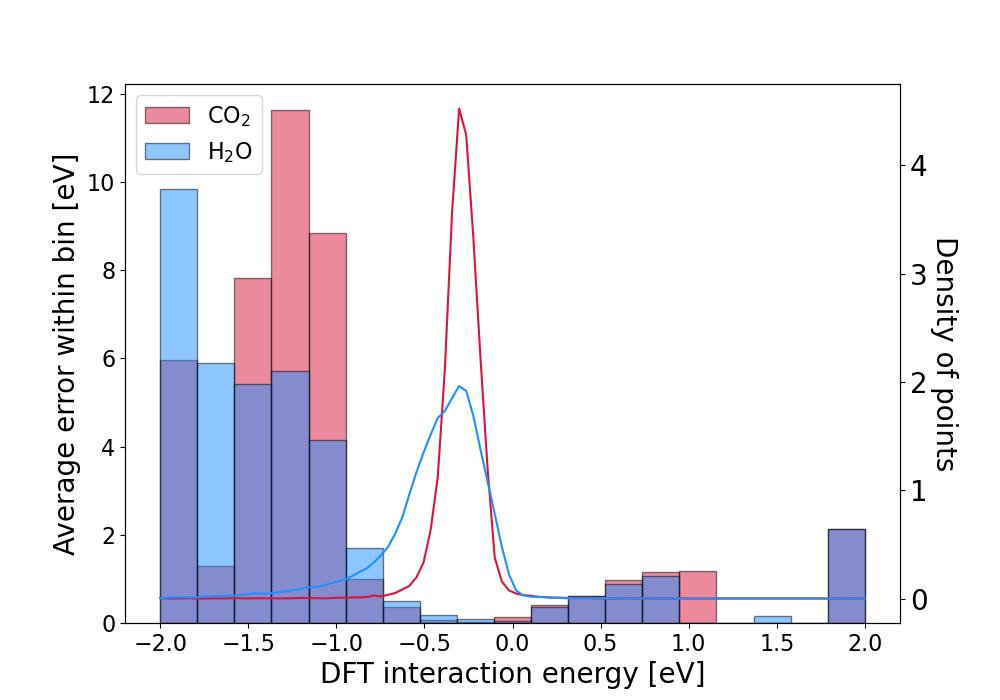}
    \caption{Binned FF errors and DFT interaction energy distributions split by adsorbate for all 51,252 systems with $-$2 $\leq$ $E_{\text{int}}^{\text{DFT}}$ $\leq 2$ eV irrespective of FF interaction energy.}
    \label{fig:Fig_FF_error_full}
\end{figure*}

\begin{figure*}[ht!]
    \includegraphics[width=\textwidth]{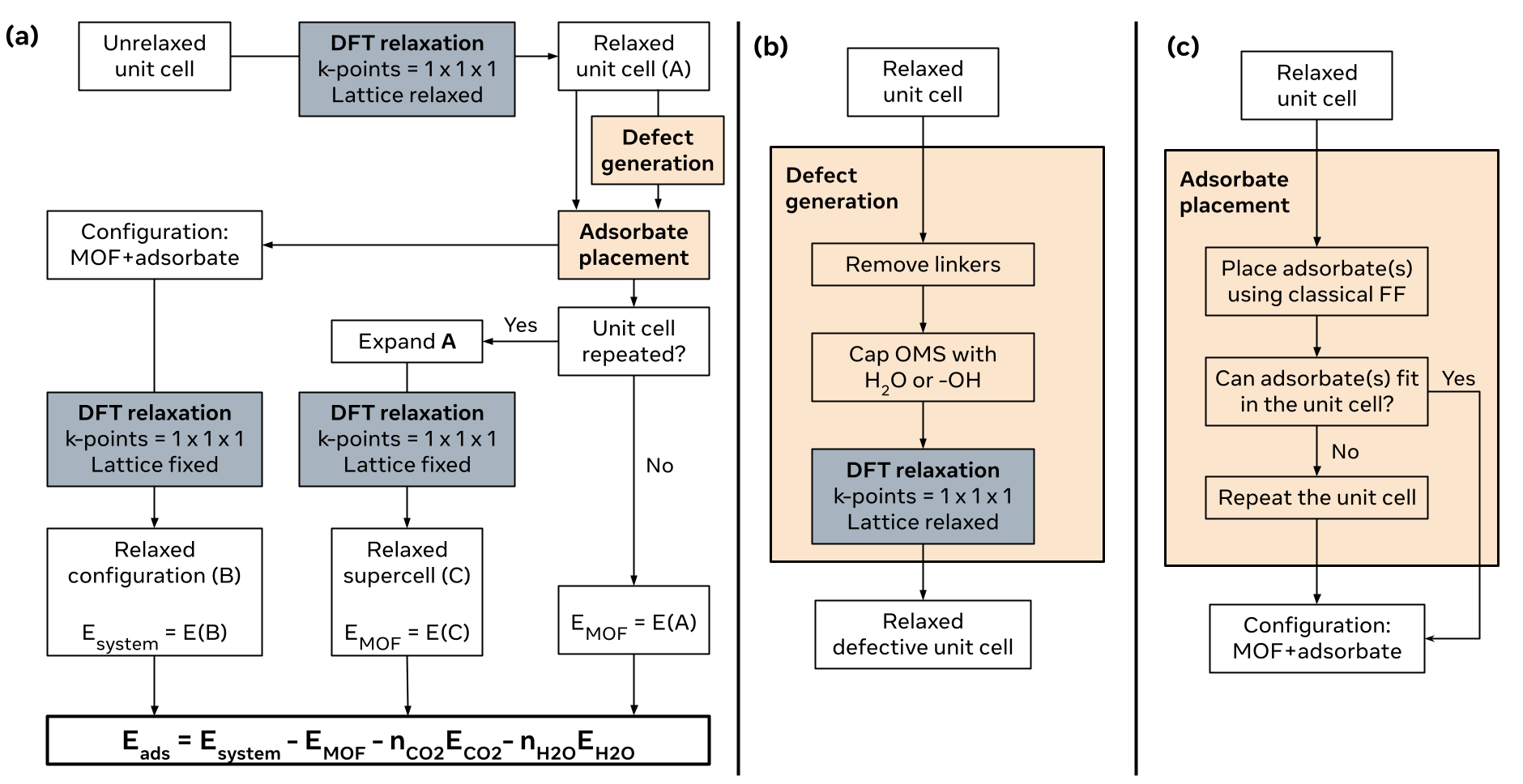}
    \caption{The workflow for generating and relaxing pristine and defective MOF structures in ODAC23.}
    \label{fig:flowchart}
\end{figure*}

\clearpage
\newpage

%%% Put all the tables here %%%
\subsection{Supplementary Tables}

\begin{table*}[h!]
    \centering
    \renewcommand{\arraystretch}{1.0}
    \setlength{\tabcolsep}{5pt}
    \renewcommand{\arraystretch}{1.0}
    \setlength{\tabcolsep}{6pt}
    \resizebox{0.97\linewidth}{!}{
    \begin{tabular}{clrrrrr}
        \toprule
        Task & Split & MOF + \carbondioxide & MOF + \water & MOF + \carbondioxide + \water & MOF + \carbondioxide + 2\water & Total\\
        \midrule
        \multirow{7}{*}{S2EF} & train & 6,608,649 & 5,196,597 & 13,092,633 & 10,973,416 & 35,871,295 \\
& val & 160,841 & 125,984 & 310,093 & 242,647 & 839,565 \\
& test-id & 163,574 & 133,372 & 360,413 & 316,156 & 973,515 \\
& test-ood (big) & 62,718 & 63,711 & 136,374 & 118,416 & 381,219 \\
& test-ood (linker) & 50,392 & 14,242 & 99,376 & 123,115 & 287,125 \\
& test-ood (topology) & 71,384 & 60,308 & 182,255 & 158,309 & 472,256 \\
& test-ood (linker \& topology) & 27,281 & 24,351 & 60,121 & 47,020 & 158,773 \\
\cmidrule{2-7}
& Total & \textbf{7,144,839} & \textbf{5,618,565} & \textbf{14,241,265} & \textbf{11,979,079} & \textbf{38,983,748} \\
        \midrule
        \multirow{7}{*}{IS2RE/IS2RS} & train & 46,274 & 34,456 & 48,373 & 33,121 & 162,224 \\
& val & 1,138 & 862 & 1,211 & 787 & 3,998 \\
& test-id & 1,291 & 972 & 1,420 & 986 & 4,669 \\
& test-ood (big) & 533 & 383 & 534 & 318 & 1,768 \\
& test-ood (linker) & 355 & 87 & 383 & 357 & 1,182 \\
& test-ood (topology) & 439 & 306 & 533 & 334 & 1,612 \\
& test-ood (linker \& topology) & 166 & 135 & 172 & 106 & 579 \\
\cmidrule{2-7}
& Total & \textbf{50,196} & \textbf{37,201} & \textbf{52,626} & \textbf{36,009} & \textbf{176,032} \\
        \bottomrule
    \end{tabular}
    }
    \caption{Summary of ODAC23 dataset organised by task and split}
    \label{tab:data_profile}
    \vspace{-10pt}
\end{table*}

\begin{table*}[h!]
    \centering
    \renewcommand{\arraystretch}{1.0}
    \setlength{\tabcolsep}{5pt}
    \renewcommand{\arraystretch}{1.0}
    \setlength{\tabcolsep}{6pt}
    \resizebox{0.85\linewidth}{!}{
    \begin{tabular}{m{8.5cm} m{4cm} c}
    \toprule
    SMILES & Structure & Number of Pristine MOFs\\
    \midrule
    \texttt{$\text{[O-]C(=O)c1ccc(cc1)C(=O)[O-]}$} & \includegraphics[scale=0.4]{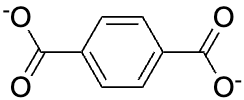} & 316\\
    \midrule
    \texttt{$\text{[O-]C(=O)c1cc(cc(c1)C(=O)[O-])C(=O)[O-]}$} & \includegraphics[scale=0.4]{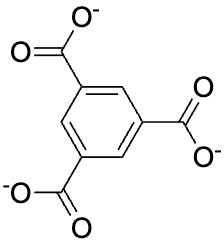} & 244\\
    \midrule
    \texttt{$\text{C\#N}$} & \includegraphics[scale=0.4]{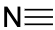} & 220\\
    \midrule
    \texttt{$\text{n1ccc(cc1)c1ccncc1}$} & \includegraphics[scale=0.4]{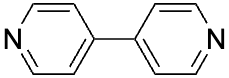} & 216\\
    \midrule
    \texttt{$\text{[O-]P(=O)([O-])[O-]}$} & \includegraphics[scale=0.4]{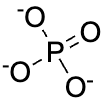} & 157\\
    \midrule
    \texttt{$\text{[O-]C(=O)C(=O)[O-]}$} & \includegraphics[scale=0.4]{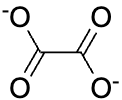} & 151\\
    \midrule
    \texttt{$\text{[O-]C=O}$} & \includegraphics[scale=0.4]{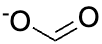} & 99\\
    \bottomrule
    \end{tabular}
    }
    \caption{Top 7 organic linkers in the pristine MOFs}
    \label{tab:top_org_linker}
    % \vspace{-10pt}
\end{table*}
\begin{table*}[h]
    \centering
    \renewcommand{\arraystretch}{1.0}
    \setlength{\tabcolsep}{5pt}
    \renewcommand{\arraystretch}{1.0}
    \setlength{\tabcolsep}{6pt}
    \resizebox{1\linewidth}{!}{
    \begin{tabular}{lccccccc}
        \toprule
        MOF & $E_{\text{ads}}$(\carbondioxide) & $E_{\text{ads}}$(\water) & $E_{\text{ads}}$(\carbondioxide) - $E_{\text{ads}}$(\water)$\uparrow$ & $E_{\text{ads}}$(\carbondioxide + \water) &  $E_{\text{inter\_mol}}^{\text{1st}}$ & $E_{\text{ads}}(\carbondioxide + 2\water)$ &  $E_{\text{inter\_mol}}^{\text{2nd}}$\\
        \midrule
 IPIDUH & -1.98 & 0.32 & -2.30 & -0.63 & 1.04 & -2.51 & -2.20 \\
KOQLUZ & -1.50 & 0.70 & -2.19 & -3.11 & -2.31 & -3.54 & -1.13 \\ 
        LEWZET & -0.89 & 0.81 & -1.69 & 0.14 & 0.23 & -4.53 & -5.48 \\ 
        TUGTAR & -1.74 & -0.47 & -1.27 & -1.70 & 0.51 & ~ & ~ \\ 
        cm503311x\_alf175K & -0.64 & 0.48 & -1.12 & -0.03 & 0.13 & -2.20 & -2.66 \\ 
        TONWUO & -1.71 & -0.63 & -1.09 & -2.68 & -0.34 & -3.78 & -0.48 \\ 
        IMAGAG & -1.14 & -0.07 & -1.07 & -1.85 & -0.64 & ~ & ~ \\ 
        EGIFUV & -1.45 & -0.41 & -1.04 & -0.43 & 1.42 & -0.87 & -0.03 \\ 
        ZIDBEV & -1.91 & -0.91 & -1.00 & -2.83 & 0.00 & -4.40 & -0.66 \\ 
        PETWIW & -0.72 & 0.28 & -0.99 & -1.38 & -0.93 & -1.09 & 0.01 \\ 
        \bottomrule
    \end{tabular}
    }
    \caption{Top 10 promising pristine MOFs with stronger \ce{CO2} adsorption energy compared to \ce{H2O} by DFT calculations [eV]}
    \label{tab:promising_pristine}
    \vspace{-10pt}
\end{table*}

\begin{table*}[h]
    \centering
    \renewcommand{\arraystretch}{1.0}
    \setlength{\tabcolsep}{5pt}
    \renewcommand{\arraystretch}{1.0}
    \setlength{\tabcolsep}{6pt}
    \resizebox{1\linewidth}{!}{
    \begin{tabular}{lcccccccc}
        \toprule
        MOF & Defect conc. & $E_{\text{ads}}$(\carbondioxide) & $E_{\text{ads}}$(\water) & $E_{\text{ads}}$(\carbondioxide)-$E_{\text{ads}}$(\water)$\uparrow$ & $E_{\text{ads}}$(\carbondioxide + \water) &  $E_{\text{inter\_mol}}^{\text{1st}}$ & $E_{\text{ads}}(\carbondioxide + 2\water)$ &  $E_{\text{inter\_mol}}^{\text{2nd}}$\\
        \midrule
  AFENEE & 0.12 & -1.86 & -0.34 & -1.52 & -0.81 & 1.40 & -1.65 & -0.50 \\ 
        OKIXIQ\_charged & 0.12 & -1.29 & -0.09 & -1.20 & -1.09 & 0.29 & -0.86 & 0.31 \\ 
        IDAGEA\_charged & 0.06 & -1.18 & -0.25 & -0.93 & -1.23 & 0.21 & -2.42 & -0.94 \\ 
        PEGCAH & 0.06 & -1.60 & -0.77 & -0.83 & -2.83 & -0.47 & -3.13 & 0.46 \\ 
        COTXEQ & 0.12 & -0.60 & 0.15 & -0.74 & -1.05 & -0.59 & -1.56 & -0.66 \\ 
        ODAHIK\_charged & 0.12 & -1.02 & -0.30 & -0.71 & -1.60 & -0.28 & -1.97 & -0.06 \\
        HUWDEJ & 0.12 & -1.36 & -0.65 & -0.71 & -1.19 & 0.82 & -2.30 & -0.46 \\
        HAJLUA & 0.16 & -0.93 & -0.26 & -0.68 & -3.83 & -2.64 & -1.95 & 2.14 \\
        HAJLOU & 0.16 & -1.02 & -0.35 & -0.67 & -1.80 & -0.43 & -4.32 & -2.16 \\
        MALRUM & 0.04 & -1.29 & -0.65 & -0.64 & -2.14 & -0.20 & -2.86 & -0.07 \\
        \bottomrule
    \end{tabular}
}
    \caption{Top 10 promising defective MOFs with stronger \ce{CO2} adsorption energy compared to \ce{H2O} by DFT calculations  [eV]}
    \label{tab:promising_defective}
    \vspace{-10pt}
\end{table*}

\begin{table*}[h!]
\centering
\renewcommand{\arraystretch}{1.0}
\setlength{\tabcolsep}{5pt}
\renewcommand{\arraystretch}{1.0}
\setlength{\tabcolsep}{6pt}
\resizebox{0.8\linewidth}{!}{
\begin{threeparttable}[t]
\begin{tabular}{lcccccc}
\multicolumn{7}{c}{List of ODAC23 Models} \\ 
\toprule
\multirow{2}{*}{Model} & \multirow{2}{*}{\# of Parameters} & \multirow{2}{*}{Equivariant Reps} & \multicolumn{4}{c}{Tasks}\\
\cmidrule{4-7}
& & & S2EF & IS2RE-direct & IS2RE-Relax & IS2RS \\
\midrule
SchNet\cite{schutt2017schnet} & 9.1M & \xmark & \cmark & *\tnote{} & * & * \\
DimeNet++\cite{klicpera2020fast} & 1.8M & \xmark & \cmark & * & * & * \\
PaiNN\cite{schutt2021equivariant} & 20.1M & \cmark & \cmark & * & * & * \\
GemNet-OC\cite{gasteiger2022graph} & 38.9M & \xmark & \cmark & \cmark & \cmark & \cmark \\
eSCN\cite{passaro2023reducing} & 51.6M & \cmark & \cmark & \cmark & \cmark & \cmark \\
EquiformerV2\cite{liao2023equiformerv2} & 31.1M & \cmark & \cmark & \cmark & \cmark & \cmark \\
EquiformerV2 (large)\cite{liao2023equiformerv2} & 153M & \cmark & \cmark & \cmark & \cmark & \cmark \\
\bottomrule
\end{tabular}
\begin{tablenotes}%\footnotesize
\item[*]Skipped because S2EF results were not competitive
\end{tablenotes}
\end{threeparttable}
}
\caption{List of ML model architectures used in this work and the different tasks they were used for}
\label{tab:model_list}
\vspace{-10pt}
\end{table*}
% predicting S2EF structure to energy and forces with different subsplits
\setlength{\tabcolsep}{3pt}
\begin{table*}[h!]
    \begin{center}
        \resizebox{0.8\linewidth}{!}{
            \begin{tabular}{p{0.14\textwidth}lcccc}
            \multicolumn{6}{c}{\gls{S2EF} Test}  \\
            \midrule
            Split & Model & \fmae & \fcos & \emae & \efwt \\
            \midrule
\multirow{8}{*}{test-id}
 &  Median baseline & 16.02 & 0.001 & 0.406 & 0.00\% \\
 &  SchNet & 14.44 & 0.254 & 0.368 & 0.02\% \\
 &  DimeNet++ & 14.31 & 0.226 & 0.416 & 0.02\% \\
 &  PaiNN & 13.04 & 0.345 & 0.241 & 0.11\% \\
 &  GemNet-OC & 9.87 & 0.605 & 0.153 & 1.16\% \\
 &  eSCN & 9.15 & 0.658 & 0.165 & 1.84\% \\
 % &  EquiformerV2 & 8.57 & 0.685 & 0.155 & 1.94\% \\
 % 11/16 updated EquiformerV2:
 &  EquiformerV2 & 7.26 & 0.674 & 0.182 & 1.97\% \\
 &  EquiformerV2 (large) & 8.20 & 0.685 & 0.145 & 2.61\% \\

\midrule
\multirow{8}{*}{test-ood(b)}
 &  Median baseline & 7.98 & 0.001 & 0.334 & 0.00\% \\
 &  SchNet & 8.16 & 0.146 & 0.529 & 0.00\% \\
 &  DimeNet++ & 7.85 & 0.157 & 0.728 & 0.00\% \\
 &  PaiNN & 7.53 & 0.242 & 0.282 & 0.12\% \\
 &  GemNet-OC & 6.19 & 0.495 & 0.207 & 0.81\% \\
 &  eSCN & 5.62 & 0.559 & 0.170 & 1.15\% \\
 % &  EquiformerV2 & 5.09 & 0.597 & 0.192 & 1.34\% \\
 % 11/16 updated EquiformerV2:
 &  EquiformerV2 & 4.91 & 0.610 & 0.166 & 1.92\% \\
 &  EquiformerV2 (large) & 4.75 & 0.612 & 0.135 & 3.07\% \\

\midrule
\multirow{8}{*}{test-ood(l)}
 &  Median baseline & 14.65 & 0.000 & 0.378 & 0.00\% \\
 &  Schnet & 13.36 & 0.262 & 0.474 & 0.00\% \\
 &  Dimenet++ & 12.15 & 0.253 & 0.501 & 0.01\% \\
 &  PaiNN & 11.47 & 0.378 & 0.252 & 0.05\% \\
 &  Gemnet-OC & 8.98 & 0.640 & 0.182 & 0.29\% \\
 &  eSCN & 7.69 & 0.719 & 0.179 & 0.59\% \\
 % &  EquiformerV2 & 7.43 & 0.751 & 0.182 & 0.54\% \\
 % 11/16 updated EquiformerV2:
 &  EquiformerV2 & 6.85 & 0.760 & 0.161 & 1.36\% \\
 &  EquiformerV2 (Large) & 6.42 & 0.761 & 0.175 & 2.03\% \\
\midrule
\multirow{8}{*}{test-ood(t)}
 &  Median baseline & 16.18 & 0.000 & 0.677 & 0.00\% \\
 &  SchNet & 14.83 & 0.181 & 1.001 & 0.00\% \\
 &  DimeNet++ & 13.62 & 0.183 & 1.297 & 0.00\% \\
 &  PaiNN & 18.18 & 0.267 & 0.507 & 0.01\% \\
 &  GemNet-OC & 12.59 & 0.488 & 0.312 & 0.05\% \\
 &  eSCN & 12.79 & 0.560 & 0.370 & 0.09\% \\
 % &  EquiformerV2 & 11.31 & 0.600 & 0.438 & 0.23\% \\
 % 11/16 updated EquiformerV2:
 &  EquiformerV2 & 10.19 & 0.617 & 0.341 & 0.52\% \\
 &  EquiformerV2 (large) & 8.80 & 0.631 & 0.292 & 0.51\% \\

\midrule
\multirow{8}{*}{test-ood(lt)}
 &  Median baseline & 13.71 & 0.000 & 0.528 & 0.00\% \\
 &  SchNet & 13.23 & 0.234 & 0.746 & 0.00\% \\
 &  DimeNet++ & 12.44 & 0.234 & 0.886 & 0.00\% \\
 &  PaiNN & 11.97 & 0.349 & 0.417 & 0.00\% \\
 &  GemNet-OC & 10.22 & 0.589 & 0.335 & 0.05\% \\
 &  eSCN & 8.78 & 0.680 & 0.305 & 0.25\% \\
 % &  EquiformerV2 & 8.23 & 0.715 & 0.404 & 0.37\% \\
 % 11/16 updated EquiformerV2:
 &  EquiformerV2 & 7.31 & 0.727 & 0.302 & 0.52\% \\
 &  EquiformerV2 (large) & 7.20 & 0.720 & 0.316 & 0.48\% \\
 \bottomrule
    \end{tabular}}
    \end{center}
    \caption{Results on the S2EF task for the various test splits}
    \label{tab:final_S2EF_subsplit_results}
\end{table*}
\setlength{\tabcolsep}{1.4pt}

\setlength{\tabcolsep}{3pt}
\begin{table*}[h!]
    \begin{center}
        \resizebox{0.8\linewidth}{!}{
            \begin{tabular}{lcccc}
            \multicolumn{5}{c}{\gls{S2EF} Test - Open Metal Sites}  \\
            \midrule
            \multirow{2}{*}{Model} & \multicolumn{2}{c}{OMS} & \multicolumn{2}{c}{Non-OMS} \\
            \cmidrule(l){2-3} \cmidrule(l){4-5}
                & \emae & \fmae & \emae & \fmae \\
            \midrule
Median baseline & 0.433 & 16.25 & 0.355 & 15.50 \\
GemNet-OC & 0.164 & 10.04 & 0.129 & 9.47 \\
eSCN & 0.186 & 9.29 & 0.120 & 8.82 \\
% EquiformerV2 & 0.163 & 8.24 & 0.117 & 8.09 \\
% 11/16 updated EquiformerV2:
EquiformerV2 & 0.204 & 8.18 & 0.169 & 6.85 \\
         \bottomrule
            \end{tabular}}
    \end{center}
    \caption{Comparison of S2EF metrics for MOFs with and without OMSs.}
    \label{tab:final_S2EF_oms_results}
\end{table*}
\setlength{\tabcolsep}{1.4pt}

\setlength{\tabcolsep}{3pt}
\begin{table*}[h!]
    \begin{center}
        \resizebox{0.8\linewidth}{!}{
            \begin{tabular}{lcccc}
            \multicolumn{5}{c}{\gls{S2EF} Test - Pristine vs Defective}  \\
            \midrule
            \multirow{2}{*}{Model} & \multicolumn{2}{c}{Pristine} & \multicolumn{2}{c}{Defective} \\
            \cmidrule(l){2-3} \cmidrule(l){4-5}
                & \emae & \fmae & \emae & \fmae \\
            \midrule
Median baseline & 0.406 & 16.02 & 0.395 & 12.12 \\
GemNet-OC & 0.153 & 9.87 & 0.182 & 8.05 \\
eSCN & 0.165 & 9.15 & 0.199 & 7.36 \\
% EquiformerV2 & 0.148 & 8.19 & 0.183 & 6.62 \\
% 11/16 updated EquiformerV2:
EquiformerV2 & 0.187 & 8.11 & 0.176 & 6.58 \\
         \bottomrule
            \end{tabular}}
    \end{center}
    \caption{Comparison of S2EF metrics for pristine and defective MOFs.}
    \label{tab:final_S2EF_defective_results}
\end{table*}
\setlength{\tabcolsep}{1.4pt}

% predicting S2EF structure to energy and forces with different subsplits
\setlength{\tabcolsep}{3pt}
\begin{table*}[h!]
    \begin{center}
        \resizebox{0.7\linewidth}{!}{
            \begin{tabular}{p{0.15\textwidth}p{0.12\textwidth}l cc}
            \multicolumn{5}{c}{\gls{IS2RE} Test}  \\
            \midrule
            Split & Method & Model & Energy MAE [eV] $\downarrow$ & EwT $\uparrow$ \\
            \midrule
\multirow{10}{*}{test-id} & \multirow{3}{*}{Direct}
 &  GemNet-OC & 0.181 & 10.40\% \\
 &  &  eSCN & 0.179 & 11.11\% \\
 &  &  EquiformerV2 & 0.172 & 10.77\% \\

\cmidrule{2-5}
& \multirow{7}{*}{Relaxation}
 &  SchNet & 0.485 & 3.10\% \\
 &  &  DimeNet++ & 0.496 & 3.23\% \\
 &  &  PaiNN & 0.225 & 9.10\% \\
 &  &  GemNet-OC & 0.174 & 12.18\% \\
 &  &  eSCN & 0.200 & 12.33\% \\
 % &  &  EquiformerV2 & 0.182 & 11.90\% \\
 % 11/16 updated EquiformerV2:
 &  &  EquiformerV2 & 0.227 & 11.59\% \\
 &  &  EquiformerV2 (large) & 0.169 & 14.47\% \\

\midrule
\multirow{10}{*}{test-ood(b)} & \multirow{3}{*}{Direct}
 &  GemNet-OC & 0.220 & 7.13\% \\
 &  &  eSCN & 0.206 & 8.31\% \\
 &  &  EquiformerV2 & 0.197 & 7.35\% \\

\cmidrule{2-5}
& \multirow{7}{*}{Relaxation}
 &  SchNet & 0.621 & 1.64\% \\
 &  &  DimeNet++ & 0.801 & 1.36\% \\
 &  &  PaiNN & 0.238 & 6.73\% \\
 &  &  GemNet-OC & 0.258 & 7.43\% \\
 &  &  eSCN & 0.289 & 9.54\% \\
 % &  &  EquiformerV2 & 0.273 & 6.28\% \\
 % 11/16 updated EquiformerV2:
 &  &  EquiformerV2 & 0.276 & 6.96\% \\
 &  &  EquiformerV2 (large) & 0.179 & 8.98\% \\

\midrule
\multirow{10}{*}{test-ood(l)} & \multirow{3}{*}{Direct}
 &  Gemnet-OC & 0.220 & 8.84\% \\
 &  &  eSCN & 0.217 & 12.50\% \\
 &  &  EquiformerV2 & 0.223 & 12.50\% \\
 
\cmidrule{2-5}
& \multirow{7}{*}{Relaxation}
 &  Schnet & 0.621 & 3.72\% \\
 &  &  Dimenet++ & 0.651 & 2.20\% \\
 &  &  PaiNN & 0.248 & 7.01\% \\
 &  &  Gemnet-OC & 0.244 & 8.19\% \\
 &  &  eSCN & 0.353 & 5.32\% \\
 % &  &  EquiformerV2 & 0.263 & 8.78\% \\
 % 11/16 updated EquiformerV2:
 &  &  EquiformerV2 & 0.241 & 9.98\% \\
 &  &  EquiformerV2 (large) & 0.232 & 10.47\% \\

\midrule
\multirow{10}{*}{test-ood(t)} & \multirow{3}{*}{Direct}
 &  GemNet-OC & 0.494 & 4.05\% \\
 &  &  eSCN & 0.404 & 4.73\% \\
 &  &  EquiformerV2 & 0.450 & 8.11\% \\

\cmidrule{2-5}
& \multirow{7}{*}{Relaxation}
 &  SchNet & 1.040 & 0.62\% \\
 &  &  DimeNet++ & 1.327 & 0.43\% \\
 &  &  PaiNN & 0.473 & 5.14\% \\
 &  &  GemNet-OC & 0.399 & 8.04\% \\
 &  &  eSCN & 0.440 & 7.23\% \\
 % &  &  EquiformerV2 & 0.375 & 8.29\% \\
 % 11/16 updated EquiformerV2:
 &  &  EquiformerV2 & 0.441 & 6.39\% \\
 &  &  EquiformerV2 (large) & 0.366 & 8.15\% \\

\midrule
\multirow{10}{*}{test-ood(lt)} & \multirow{3}{*}{Direct}
 &  GemNet-OC & 0.385 & 6.68\% \\
 &  &  eSCN & 0.360 & 7.02\% \\
 &  &  EquiformerV2 & 0.336 & 6.51\% \\

\cmidrule{2-5}
& \multirow{7}{*}{Relaxation}
 &  SchNet & 0.711 & 2.05\% \\
 &  &  DimeNet++ & 1.116 & 0.68\% \\ &  &  PaiNN & 0.410 & 5.14\% \\
 &  &  GemNet-OC & 0.397 & 8.45\% \\
 &  &  eSCN & 0.463 & 5.10\% \\
 % &  &  EquiformerV2 & 0.466 & 8.05\% \\
 % 11/16 updated EquiformerV2:
 &  &  EquiformerV2 & 0.414 & 6.56\% \\
 &  &  EquiformerV2 (large) & 0.405 & 9.87\% \\
\bottomrule
            \end{tabular}
            }
    \end{center}
    \caption{Full results of metrics for IS2RE task on all data splits.}
    \label{tab:final_IS2RE_subsplit_results}
\end{table*}
\setlength{\tabcolsep}{1.4pt}
\setlength{\tabcolsep}{3pt}
\begin{table*}[h!]
\begin{center}
\resizebox{0.65\linewidth}{!}{
\begin{tabular}{p{0.2\textwidth}lccc}
\multicolumn{5}{c}{\gls{IS2RS} Test}  \\
\midrule
Split & Model & ADwT $\uparrow$ & FbT $\uparrow$ & AFbT $\uparrow$ \\
\midrule
\multirow{4}{*}{ test-id }  &  GemNet-OC & 85.46\% & 0.00\% & 6.53\% \\
&  eSCN & 85.13\% & 0.40\% & 11.32\% \\
% &  EquiformerV2 & 84.50\% & 0.60\% & 11.77\% \\
% 11/16 updated EquiformerV2:
&  EquiformerV2 & 87.92\% & 0.00\% & 12.05\% \\
&  EquiformerV2 (large) & 87.37\% & 0.60\% & 11.44\% \\
\midrule
\multirow{4}{*}{ test-ood(b) }  &  GemNet-OC & 87.79\% & 0.00\% & 4.54\% \\
&  eSCN & 86.30\% & 0.60\% & 3.41\% \\
% &  EquiformerV2 & 83.17\% & 0.40\% & 4.11\% \\
% 11/16 updated EquiformerV2:
&  EquiformerV2 & 88.57\% & 0.00\% & 3.28\% \\
&  EquiformerV2 (large) & 87.50\% & 0.40\% & 4.50\% \\
\midrule
\multirow{4}{*}{ test-ood(l) }  &  GemNet-OC & 69.74\% & 0.00\% & 1.97\% \\
&  eSCN & 66.56\% & 0.40\% & 4.42\% \\
% &  EquiformerV2 & 64.32\% & 0.20\% & 4.47\% \\
% 11/16 updated EquiformerV2:
&  EquiformerV2 & 74.34\% & 0.00\% & 4.83\% \\
&  EquiformerV2 (large) & 75.44\% & 0.00\% & 4.78\% \\
\midrule
\multirow{4}{*}{ test-ood(t) }  &  GemNet-OC & 60.03\% & 0.00\% & 0.95\% \\
&  eSCN & 60.08\% & 0.00\% & 1.89\% \\
% &  EquiformerV2 & 52.50\% & 0.00\% & 1.93\% \\
% 11/16 updated EquiformerV2:
&  EquiformerV2 & 68.23\% & 0.00\% & 1.58\% \\
&  EquiformerV2 (large) & 66.97\% & 0.20\% & 2.50\% \\
\midrule
\multirow{4}{*}{ test-ood(lt) }  &  GemNet-OC & 59.11\% & 0.00\% & 1.64\% \\
&  eSCN & 61.27\% & 0.00\% & 3.54\% \\
% &  EquiformerV2 & 59.64\% & 0.00\% & 3.15\% \\
% 11/16 updated EquiformerV2:
&  EquiformerV2 & 68.30\% & 0.00\% & 3.27\% \\
&  EquiformerV2 (large) & 65.58\% & 0.20\% & 2.32\% \\
\bottomrule
\end{tabular}
}
\end{center}
\caption{Full results of metrics for IS2RS task on all data splits.}
\label{tab:final_IS2RS_subsplit_results}
\end{table*}
\setlength{\tabcolsep}{1.4pt}

\clearpage

\end{document}